\documentclass[twocolumn,showpacs,longbibliography]{revtex4-1}

\usepackage[dvips]{graphicx}
\usepackage{amsmath}
\usepackage{amssymb}
\usepackage{subfigure}
\usepackage{layout}
\usepackage{comment}

\usepackage{array}
\usepackage{color}
\newcolumntype{M}{>{\centering\arraybackslash}m{2.5cm}}
\newcolumntype{S}{>{\centering\arraybackslash}m{1.5cm}}
\usepackage[normalem]{ulem}

\newcommand{\ve}[1]{\mathbf{#1}}
\newcommand{\refeq}[1]{(\ref{#1})}

\begin{document}

\title{The role of volume and surface spontaneous parametric down-conversion in the generation of photon pairs in layered media}

\author{D. Jav\r{u}rek}
\address{RCPTM, Joint Laboratory of Optics of Palack\'y University
and Institute of Physics of CAS, 17. listopadu 12, 771 46 Olomouc,
Czech Republic} \email{javurek@slo.upol.cz}

\author{J. Pe\v{r}ina Jr.}
\address{Institute of Physics, Joint Laboratory of Optics of
Palack\'y University and Institute of Physics of CAS, 17.
listopadu 50a, 771 46 Olomouc, Czech Republic}

\begin{abstract}
A rigorous description of volume and surface spontaneous
parametric down-conversion in 1D nonlinear layered structures is
developed considering exact continuity relations for the fields'
amplitudes at the boundaries. The nonlinear process is described
by the quantum momentum operator that provides the Heisenberg
equations which solution is continuous at the boundaries. The
transfer-matrix formalism is applied. The volume and surface
contributions are clearly identified. Numerical analysis of a
structure composed of 20 alternating GaN/AlN layers is given as an
example.
\end{abstract}



\maketitle

\section{Introduction}

Spontaneous parametric down-conversion (SPDC) is a second-order
nonlinear process \cite{Louisell1961,Harris1967,Magde1967} in
which one photon with higher energy is annihilated and two photons
of lower energies are simultaneously created. Due to the laws of
energy and momentum conservations quantum correlations
(entanglement) between the photons in a pair emerge
\cite{Keller1997,Svozilik2012,Grice2008,Javurek2014b}. The process
of SPDC occurs either inside the media with non-zero second-order
permittivity tensor (non-centrosymmetric crystals) or at the
boundaries of these media
\cite{PerinaJr2009,PerinaJr2014,Boyd2003}.

The process of SPDC has been observed in nonlinear bulk media
\cite{Boyd2003,Dmitriev1999}, systems of nonlinear thin layers
\cite{PerinaJr2006,PerinaJr2009c,PerinaJr2011} including
metallo-dielectric layers
\cite{Javurek2014b,Javurek2012,Javurek2014x}, nonlinear photonic
fibers \cite{Zhu2012,Javurek2014a,Javurek2014b}, nonlinear
photonic waveguides
\cite{Eckstein2011,Jachura2014,Machulka2013,Clausen2014}, as well
as in complex nonlinear photonic structures \cite{Chen2014}. Bulk
media including the most common nonlinear crystals LiNbO$_3$ and
KTP represent the historically oldest sources of photon pairs.
Periodically-poled nonlinear crystals with their freedom in
tailoring phase-matching conditions have been obtained later, at
the beginning of the 90's \cite{Hayata1991}. At the same time
periodically poled waveguides \cite{Lim1989,Shinozaki1992} and
fibers \cite{Kashyap1991,Chmela1991} followed them up.

In the process of SPDC in homogeneous bulk media, phase matching
of all three interacting fields (pump, signal and idler) in the
direction of their propagation as well as in their transverse
planes is needed to arrive at efficient nonlinear interaction.
Phase-matching conditions can be achieved by angle or temperature
tuning of birefringent crystals \cite{Boyd2003}. Or,
alternatively, by poling a nonlinear crystal
\cite{Harris2007,Brida2009,Svozilik2009,Svozilik2011a} which
results in quasi-phase-matching conditions. However, if the length
of a nonlinear medium is comparable to the interacting fields'
wavelengths, the phase-matching condition does not play an
important role in reaching efficient nonlinear interaction.
Instead, the overlap of electric-field amplitudes of the
interacting fields inside the medium is crucial. Moreover, the
contribution of the nonlinear interaction around the boundaries of
such thin media becomes important
\cite{Blombergen1962,Blombergen1969,Mlejnek1999,Centini2008}.
Provided that the number of boundaries per unit length (or volume)
of the crystal (photonic structure) is sufficiently high, the
emission rate of photon pairs coming from the boundaries may even
be comparable to the emission rate of photon pairs created in the
volume \cite{PerinaJr2009a}. This also concerns the nonlinear
poled structures in which the boundaries are formed inbetween the
domains with different signs of $ \chi^{(2)} $ susceptibility.

Theoretical approaches to SPDC in layered structures (including
poled crystals) have been developed in the Schr\"{o}dinger as well
as the Heisenberg pictures. In the Schr\"{o}dinger picture, a
perturbation solution of the Schr\"{o}dinger equation in the
nonlinear coupling constant was found. In the first order, it
describes the generation of one photon pair
\cite{PerinaJr2011,Javurek2012,Javurek2014c}. On the other hand,
linear Heisenberg equations occur in the Heisenberg picture. They
allow to treat the nonlinear interaction for arbitrarily intense
signal and idler fields. Their solution can also be conveniently
written such that the continuity requirements of the electric- and
magnetic-field operator amplitudes at the boundaries are
fulfilled. This allows to describe simultaneously the volume and
surface contributions to SPCD.

Using the Heisenberg picture, the volume contribution to SPDC has
been widely studied in
Refs.~\cite{Huttner1990,BenAryeh1991,PerinaJr2011} whereas the
emission of photon pairs at the boundaries has been treated in
Ref.~\cite{PerinaJr2009a} applying the perturbation technique. The
perturbation approach allowed to introduce corrections to the
creation and annihilation operators of the signal and idler fields
independently and then to apply the transfer-matrix formalism.
Contrary to this, the theory developed here treats the fields at
the boundaries in general which results in the coupling between
the signal- and idler-field operators analyzed first in
Ref.~\cite{Luks2012,Perinova2013} for the cw-interaction. This
means that the developed theory is more general (and more precise)
compared to that of Ref.~\cite{PerinaJr2009a}, though it requires
an extensive numeric approach. Moreover, it clearly identifies the
fields arising in the volume and surface SPDC.

The paper is structured as follows. In Sec.~II, the model
describing both volume and surface SPDC is developed. In Sec.~III,
quantities characterizing photon pairs and derived from the
general solution are defined. Results of numerical simulations are
discussed in Sec.~IV. Conclusions are drawn in Sec.~V.

\section{Volume and surface spontaneous parametric down-conversion}

\begin{figure} 
 \includegraphics[scale=1]{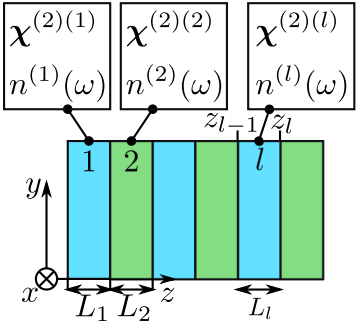}
 \caption{[Color online] Scheme of a layered structure. Second-order tensor $\boldsymbol{\chi}^{(2)(l)}$
  of nonlinear susceptibility characterizes an $l$-th layer with
  index $n^{(l)}(\omega)$ of refraction; $L_l$ is the length of the $l$-th layer and $z_l$ ($z_{l+1}$)
  denotes the position of its left (right) boundary.}
 \label{f1}
\end{figure}

The proposed model of SPDC is appropriate for 1D nonlinear
photonic structures composed of parallel layers (or domains)
having in general different material parameters and lengths. As an
example, we consider a layered structure composed of alternating
layers with different linear indices of refraction and nonlinear
susceptibilities (see Fig.~\ref{f1}).

The process of SPCD is assumed to be pumped by a strong
(un-depleted) classical field. In a layered structure, the
positive-frequency vectorial electric-field amplitude
$\mathbf{E}_{\rm p}^{(+)}(z,t)$ of the pump beam can conveniently
be decomposed as follows \cite{PerinaJr2011,PerinaJr2009a}:
\begin{align} 
\nonumber
 \mathbf{E}_{\rm p}^{(+)}(z,t) =& \int_{0}^{\infty} d\omega_{\rm p} \sum_{l=0}^{N+1} \mathrm{rect}^{(l)}(z)
  \sum_{g={\rm F,B}}^{\gamma={\rm x,y}} \ve{e}_{{\rm p},\gamma} A_{{\rm p}_g,\gamma}^{(l)}(\omega_{\rm p}) \\
  & \times \mathrm{exp}\left[ik^{(l)}_{{\rm p}_g,\gamma}(\omega_{\rm p})(z-z_{l})
  -i\omega_{\rm p} t\right];
  \label{e1}
\end{align}
$\omega_{\rm p} $ denotes the angular frequency of the pump beam.
Function $\mathrm{rect}^{(l)}(z)$ is nonzero only for $
z\in(z_{l},z_{l+1}) $ where $\mathrm{rect}^{(l)}(z)=1 $,
$l\in\{1,\ldots,N\}$. For the input [output] medium, we have
$\mathrm{rect}^{(0)}(z)=1 $ for $ z\in(-\infty,z_1) $ [$
\mathrm{rect}^{(N+1)}(z)=1 $ for $ z\in(z_{N+1},\infty)$] and it
equals zero otherwise. Amplitude $A_{{\rm p}_g,\gamma}^{(l)}$
occurring in Eq.~(\ref{e1}) denotes the spectral pump
electric-field amplitude at the left boundary of an $l$-th layer.
The forward- (backward-) propagating fields are indicated by index
$ {\rm F} $ ($ {\rm B} $). The unit electric-field vectors
$\ve{e}_{{\rm p},\gamma}$ determine the field's polarization
either along the $ {\rm x} $ or $ {\rm  y} $ axis
\cite{Yeh1988,PerinaJr2011,PerinaJr2014}. The wave numbers
$k_{{\rm p}_g,\gamma}^{(l)}$ satisfy the linear dispersion
relations appropriate to an $l$-th layer, $k^{(l)}_{{\rm
p}_g,\gamma}(\omega_{\rm p})=\pm(\omega_{\rm p}/c)n_{{\rm
p},\gamma}^{(l)}(\omega_{\rm p})$, $c$ being the speed of light in
vacuum and $n_{{\rm p},\gamma}^{(l)}(\omega_{\rm p})$ denoting
index of refraction in this layer. The plus (minus) sign in the
definition of $k^{(l)}_{{\rm p}_g,\gamma}$ refers to the forward-
(backward-) propagating field. Symbol $ \sum_{g}^{\gamma} $ stands
for the summation over both the direction of field's propagation
and polarization.

The signal and idler positive-frequency vectorial electric-field
operator amplitudes $\hat{\ve{E}}^{(+)}_{\rm s}(z,t)$ and
$\hat{\ve{E}}^{(+)}_{\rm i}(z,t)$, respectively, are defined
similarly as the pump amplitude:
\begin{eqnarray} 
  \nonumber
   \hat{\ve{E}}^{(+)}_m(z,t) &=& i\int_0^{\infty} d\omega_m  \sum_{l=0}^{N+1} \tau_{m,\alpha}^{(l)}(\omega_m)
   \mathrm{rect}^{(l)}(z) \sum_{a={\rm F,B}}^{\alpha={\rm x,y}}\\
   && \hspace{-20mm} \ve{e}_{m,\alpha} \hat{a}^{(l)}_{m_a,\alpha}(z,\omega_m) \mathrm{exp}\left(-i\omega_m t\right);
   \hspace{5mm} m\in\{{\rm s,i}\}.
   \label{e2}
\end{eqnarray}
In Eq.~(\ref{e2}), amplitude $\tau_{m,\alpha}^{(l)}(\omega_m)$ per
one photon is defined as
\begin{equation} 
 \tau^{(l)}_{m,\alpha}(\omega_m) = \sqrt{\frac{\hbar \omega_m}{4\pi\varepsilon_0 c n_{m,\alpha}^{(l)}(\omega_m)A}}
 \label{e3}
\end{equation}
assuming homogeneous fields localized in transverse area $A$.
Symbols $\ve{e}_{m,\alpha}$ introduced in Eq.~(\ref{e3}) denote
the unit polarization vectors in field $ m $ and $ \hbar $ stands
for the reduced Planck constant. Operator
$\hat{a}^{(l)}_{m_a,\alpha}(z,\omega_m)$ annihilates one photon at
position $z$ in field $m$ propagating in direction $a$ with
polarization $\alpha$ and frequency $\omega_m$. The annihilation
operator $\hat{a}^{(l)}_{m_a,\alpha}(z,\omega_m)$ is assumed to
fulfill the equal-space commutation relations together with its
hermitian conjugated creation operator
$\hat{a}^{(l)\dagger}_{m_a,\alpha}(z,\omega_m)$:
\cite{Huttner1990,BenAryeh1991,PerinaJr2009a}
  \begin{eqnarray} 
  \nonumber
   [\hat{a}^{(l)}_{m_a,\alpha}(z,\omega_m),\hat{a}^{(l)\dagger}_{m'_{a'},\alpha'}(z,\omega_m')] &=& \delta_{aa'}\delta_{mm'} \delta_{\alpha \alpha'} \\
   && \times \delta(\omega_m-\omega_m').
   \label{e4}
\end{eqnarray}

Spatial evolution of the operator
$\hat{a}^{(l)}_{m_a,\alpha}(z,\omega_m)$ is given by the
Heisenberg equation
\begin{equation} 
 \frac{\partial \hat{a}^{(l)}_{m_a,\alpha}}{\partial z}(z,\omega_m) = \frac{1}{i\hbar} [\hat{G}_z(z), \hat{a}^{(l)}_{m_a,\alpha}(z,\omega_m)]
 \label{e5}
\end{equation}
derived from the following momentum operator $\hat{G}_z$
\cite{BenAryeh1991}:
\begin{eqnarray} 
\label{e6}
 \hat{G}_z(z) &=& A \hspace{-2mm}\int_{-\infty}^{\infty} dt\, \hat{\sigma}_{\rm zz}^{\mathrm{eff}}(z,t) , \\
 \nonumber
 \hat{\sigma}_{\rm zz}^{\mathrm{eff}}(z,t) &=& \sum_{m={\rm s,i}} \sum_{a={\rm F,B}}^{\alpha={\rm x,y}}
  \Bigl\{ \varepsilon_0 \hat{\ve{E}}_{m_a,\alpha}^{(-)}(z,t)\cdot\hat{\ve{E}}_{m_a,\alpha}^{(+)}(z,t) \\
 \nonumber
 &&\hspace{-20mm} + \frac{1}{\mu_0}\hat{\ve{B}}_{m_a,\alpha}^{(-)}(z,t)\cdot\hat{\ve{B}}_{m_a,\alpha}^{(+)}(z,t) \\
 \nonumber
 && \hspace{-20mm} + \varepsilon_0 \int_0^{\infty} d\omega_m \int_0^{\infty} d\omega_m'\, \chi^{(1)}(\omega_m)\\
 \nonumber
 && \hspace{-20mm} \Bigl. \times \hat{\ve{E}}_{m_a,\alpha}^{(-)}(z,\omega_m') \cdot\hat{\ve{E}}_{m_a,\alpha}^{(+)}(z,\omega_m)
  \mathrm{exp}[-i(\omega_m-\omega_m')t] \Bigr \}
 \\
 \nonumber
 && \hspace{-20mm} + 2\varepsilon_0 \int_0^{\infty} d\omega_{\rm p} \int_0^{\infty} d\omega_{\rm s} \int_0^{\infty} d\omega_{\rm i}\,
  \boldsymbol{\chi}^{(2)}(\omega_{\rm p};\omega_{\rm s},\omega_{\rm i})\mathbf{:} \ve{E}_{\rm p}(z,\omega_{\rm p})  \\ \nonumber
 \nonumber
 && \hspace{-20mm} \times \hat{\ve{E}}_{\rm s}^{\dagger}(z,\omega_{\rm s}) \hat{\ve{E}}_{\rm i}^{\dagger}(z,\omega_{\rm i})
  \mathrm{exp}\left[-i(\omega_{\rm p} - \omega_{\rm s} - \omega_{\rm i}) t\right].
  \nonumber \\
 &&
 \label{e7}
\end{eqnarray}
In Eq.~(\ref{e7}), $\hat{\sigma}_{\rm zz}^{\mathrm{eff}}$ means
the $ {\rm zz} $ component of effective Maxwell stress-tensor
operator and $\hat{\ve{E}}^{(+)}_{m_a,\alpha}(z,\omega_a)$ stands
for the spectral positive-frequency electric-field operator
amplitude [$\hat{\ve{E}}^{(+)}_m(z,t)= \sum_{a={\rm
F,B}}^{\alpha={\rm x,y}} \int_{0}^{\infty} d\omega_a\,
\hat{\ve{E}}^{(+)}_{m_a,\alpha}(z,\omega_a)\,
\mathrm{exp}(-i\omega_a t)$]. Spectral positive-frequency
magnetic-field operator amplitudes
$\hat{\ve{B}}^{(+)}_{m_a,\alpha}(z,\omega_m)$ are derived from the
Maxwell equations [$\hat{\ve{B}}^{(+)}_{m_a,{\rm y}}(z,\omega_m) =
\ve{k}_{m,\alpha}(\omega_m) \times \hat{\ve{E}}^{(+)}_{m_{a},{\rm
x}}(z,\omega) / \omega_m$]. Symbol $ \cdot $ denotes scalar product
and  operation $\boldsymbol{:}$ shorthands tensor
$\boldsymbol{\chi}^{(2)}$ with respect to its three indices.

Applying Eqs.~\refeq{e1}, \refeq{e2}, \refeq{e6}, and \refeq{e7}
the following explicit form of momentum operator $\hat{G}_{\rm
z}(z)$ is obtained:
\begin{align} 
\nonumber
 &\hat{G}_{\rm z}(z) = \sum_{m={\rm s,i}}\sum_{l=0}^{N+1} \int_0^{\infty} d\omega_m \sum_{a={\rm F,B}}^{\alpha={\rm x,y}}
  \mathrm{rect}^{(l)}(z) \hbar k^{(l)}_{m,\alpha}(\omega_m) \\
 \nonumber
 & \hspace{0mm}\times \hat{a}_{m_{a},\alpha}^{(l) \dagger}(z,\omega_m) \hat{a}^{(l)}_{m_{a},\alpha}(z,\omega_m)
   \\ \nonumber
 & \hspace{0mm} - i\hbar \int_0^{\infty} d\omega_{\rm s} \int_0^{\infty} d\omega_{\rm
  i} \sum_{a,b,g={\rm F,B}}^{\alpha,\beta,\gamma={\rm x,y}}\sum_{l=1}^{N}  T_{g}^{\alpha \beta \gamma,(l)*}(\omega_{\rm s},\omega_{\rm i})\\
  \nonumber
  & \hspace{0mm} \times \exp\left[ik_{{\rm p}_\gamma,g}^{(l)}(\omega_{\rm s}+\omega_{\rm i})(z-z_l)\right] \hat{a}^{(l) \dagger}_{{\rm s}_a,\alpha}(z,\omega_{\rm s})
   \hat{a}^{(l) \dagger}_{{\rm i}_b,\beta} (z,\omega_{\rm i}) \\
  &
  \label{e8}
\end{align}
and
\begin{align} 
\nonumber
  &T^{\alpha \beta \gamma,(l)}_{g}(\omega_{\rm s}, \omega_{\rm i}) \equiv  \frac{4 i \pi \varepsilon_0 A}{\hbar}   \tau_{{\rm s},\alpha}^{(l)}(\omega_{\rm s})
   \tau_{{\rm i},\beta}^{(l)}(\omega_{\rm i})  \\
  \nonumber
  &  \hspace{0mm}\times\boldsymbol{\chi}^{(2)(l)}(\omega_{\rm s}+\omega_{\rm i};\omega_{\rm s},\omega_{\rm i}) \mathbf{:} \ve{e}_{{\rm p},\gamma}
   \ve{e}_{{\rm s},\alpha} \ve{e}_{{\rm i},\beta} A^{(l)*}_{{\rm p}_g,\gamma}(\omega_{\rm s}+\omega_{\rm i}). \\
  &\label{e9}
\end{align}
Symbol $k_{m,\alpha}^{(l)}$ occurring in Eq.~(\ref{e8}) denotes
the absolute value of wave vector $k_{m_a,\alpha}^{(l)}$. Below,
we utilize the formalism that does not distinguish explicitly
between the forward- and backward-propagating modes via the sign
of wave vectors. Instead, appropriate signs are added to the
linear and nonlinear terms in momentum operator $\hat{G}_{\rm z}$.

Applying the commutation relations \refeq{e4} the Heisenberg
equations \refeq{e5} are obtained in their explicit form:
\begin{eqnarray}  
\nonumber
 \frac{\partial \hat{a}_{{\rm s}_a,\alpha}^{(l)}}{\partial z}(z,\omega_{\rm s}) &=& i k_{{\rm s}_a,\alpha}^{(l)} \hat{a}_{{\rm s}_a,\alpha}^{(l)}(z,\omega_{\rm s})
 + [\pm 1]_a \int_{0}^{\infty} d\omega_{\rm i}\, \\
 \nonumber
 && \hspace{-30mm}  \sum_{b,g={\rm F,B}}^{\beta,\gamma={\rm x,y}}  T_{g}^{\alpha\beta\gamma,(l)*}(\omega_{\rm s},\omega_{\rm i}) \exp[ik_{{\rm p}_g,\gamma}^{(l)}(z-z_l)]
   \hat{a}_{{\rm i}_b,\beta}^{(l),\dagger}(z,\omega_{\rm i}).
  \\
 && \label{e10b}
\end{eqnarray}
Symbol $[\pm 1]_a$ equals $+1$ for a forward propagating field
($a={\rm F}$) and $-1$ for a backward propagating field ($a={\rm
B}$).

The solution of Heisenberg equation \refeq{e10b} for the
signal-field operator $\hat{a}_{{\rm s}_a,\alpha}^{(l)}$ consists
of the homogeneous and particular solutions:
\begin{eqnarray} 
\nonumber
 \hat{a}_{{\rm s}_a,\alpha}^{(l)}(z,\omega_{\rm s}) &=& \hat{\bar{a}}^{(l)}_{{\rm s}_a,\alpha}(z_l,\omega_{\rm s}) \exp[ik_{{\rm s}_a,\alpha}^{(l)}(\omega_{\rm s})(z-z_l)] \\
 \nonumber
 &&\hspace{-20mm}+ \int_0^{\infty} d\omega_{\rm i} \sum_{b={\rm F,B}}^{\beta={\rm x,y}}\Phi_{s,ab}^{\alpha \beta,(l)*}(z,\omega_{\rm s},\omega_{\rm i})\,
  \hat{\bar{a}}_{{\rm i}_b,\beta}^{(l) \dagger}(z_l,\omega_{\rm i})
  \\
  \label{e11}
  &&\hspace{-20mm} \times \exp[ik_{{\rm s}_a,\alpha}^{(l)}(\omega_{\rm s})(z-z_a^{(l)})]
\end{eqnarray}
and
\begin{eqnarray}  
 \nonumber
 \Phi_{s,ab}^{\alpha \beta, (l)}(z,\omega_{\rm s}, \omega_{\rm i})&\equiv& i [\pm 1]_a  \sum_{g={\rm F,B}}^{\gamma={\rm x,y}}
  T_{g}^{\alpha \beta \gamma, (l)}(\omega_{\rm s}, \omega_{\rm i}) \\ \nonumber
  &&\hspace{-20mm}\times \mathrm{rect}^{(l)}(z) \exp[-i \phi_{{\rm s}_a,bg}^{\beta \gamma,(l)}(\omega_{\rm s},\omega_{\rm i})] \\
  \label{e13}
  &&\hspace{-20mm}\times \frac{\exp \left[- i \Delta k_{abg}^{\alpha \beta \gamma,(l)}(z-z_a^{(l)})\right] - 1}{\Delta k_{abg}^{\alpha \beta
  \gamma,(l)}}.
\end{eqnarray}
We note that the particular solution has been derived by the
convolution of the Green function of Eq.~\refeq{e10b} and the
nonlinear source term on the right-hand side of Eq.~\refeq{e10b}.
The signal-field annihilation operator $\hat{\bar{a}}_{{\rm
s}_a,\alpha}^{(l)}(z_l,\omega_{\rm s})$ and idler-field creation
operator $\hat{\bar{a}}_{{\rm
i}_b,\beta}^{(l)\dagger}(z_l,\omega_{\rm i})$ occurring at the
right-hand side of Eq.~(\ref{e11}) are appropriate for the
homogeneous solution and so they describe the free-field
propagation. Spatial dependence of the signal-field operator
$\hat{\bar{a}}_{{\rm s}_a,\alpha}^{(l)}(z,\omega_{\rm s})$,
considered as an example, is thus described as
\begin{equation}   
 \hat{\bar{a}}_{{\rm s}_a,\alpha}^{(l)}(z,\omega_{\rm s}) = \hat{\bar{a}}_{{\rm s}_a,\alpha}^{(l)}(z_l,\omega_{\rm s})
  \exp[i k_{{\rm s}_a,\alpha}^{(l)}(z-z_l)].
 \label{e14}
\end{equation}
The signal- and idler-field operators $\hat{\bar{a}}_{{\rm
s}_a,\alpha}^{(l)}(z,\omega_{\rm s})$ and $\hat{\bar{a}}_{{\rm
i}_b,\beta}^{(l)\dagger}(z,\omega_{\rm i})$ also obey the equal
space commutation relations
\begin{eqnarray}   
 \nonumber
 [\hat{\bar{a}}_{m_a,\alpha}^{(l)}(z,\omega_m),\hat{\bar{a}}_{{m'}_{a'},\alpha'}^{(l)}(z,\omega_{m'}')] &=& \delta_{mm'} \delta_{aa'} \delta_{\alpha \alpha'} \\
 && \hspace{-5mm} \times  \delta(\omega_m - \omega_m').
\end{eqnarray}
In Eq.~(\ref{e13}), the difference $\Delta k_{abg}^{\alpha \beta
\gamma,(l)}$ of wave vectors in an $ l $-th layer equals $\Delta
k_{abg}^{\alpha \beta \gamma,(l)} = k_{{\rm p}_g,\gamma}^{(l)} -
k_{{\rm s}_a,\alpha}^{(l)} - k_{{\rm i}_b,\beta}^{(l)}$. Position
$z_a^{(l)}$ in the $ l $-th layer equals $z_{l}$ ($z_{l+1}$) for
forward- (backward-) propagating fields. Similarly, phase factor
$\phi_{{\rm s}_a,bg}^{\beta\gamma,(l)}$ introduced in
Eq.~(\ref{e13}) is equal to zero [$ k_{{\rm
p}_g,\gamma}^{(l)}-k_{{\rm i}_b,\beta}^{(l)}) L^{(l)}$] for
forward- [backward-] propagating fields. The solution for the
idler-field operators is derived from Eq.~\refeq{e11} invoking the
symmetry between the signal and idler fields ($s \leftrightarrow
i$).

The spatial dependence of signal [idler] electric-field operator
amplitude $\hat{\ve{E}}_{\rm s}(z,t)$ [$\hat{\ve{E}}_{\rm
i}(z,t)$] inside the structure is determined once we know the
transformations between the signal [idler] operators
$\hat{\bar{a}}_{{\rm s}_a,\alpha}^{(l-1)}$ [$\hat{\bar{a}}_{{\rm
i}_b,\beta}^{(l-1)\dagger}$] and $\hat{\bar{a}}_{{\rm
s}_a,\alpha}^{(l)}$ [$\hat{\bar{a}}_{{\rm
i}_b,\beta}^{(l)\dagger}$] in all adjacent layers $l-1$ and $l$.
The transformation is derived from the boundary conditions for the
electric- and magnetic-field operators and the propagation
formula~\refeq{e14}. The boundary conditions for the signal-field
operators between layers $l-1$ and $l$ require the continuity of
electric- [$\hat{\ve{E}}_{\rm s}^{(l-1)}(z,t)$ and
$\hat{\ve{E}}_{\rm s}^{(l)}(z,t)$] and magnetic-field
[$\hat{\ve{H}}_{\rm s}^{(l-1)}(z,t)$ and $\hat{\ve{H}}_{\rm
s}^{(l)}(z,t)$] vectorial operator amplitudes. Applying the
field's decomposition written in Eq.~\refeq{e2} the boundary
conditions are transformed into the following relations:
\begin{eqnarray} 
 \nonumber
 \tau^{(l-1)}_{m,\alpha}(\omega_m) \sum_{a={\rm F,B}} \hat{a}^{(l-1)}_{m_a,\alpha}(z_l,\omega_m) &=&  \\
 \label{e31}
 && \hspace{-40mm} \tau^{(l)}_{m,\alpha}(\omega_m) \sum_{a={\rm F,B}} \hat{a}^{(l)}_{m_a,\alpha}(z_l,\omega_m), \\
 \nonumber
 \tau^{(l-1)}_{m,\alpha}(\omega_m) \sum_{a={\rm F,B}} \frac{\partial \hat{a}^{(l-1)}_{m_a,\alpha}}{\partial z}(z_l,\omega_m) &=& \\
 \nonumber
 && \hspace{-40mm} \tau^{(l)}_{m,\alpha}(\omega_m) \sum_{a={\rm F,B}} \frac{\partial \hat{a}^{(l)}_{m_a,\alpha}}{\partial z}(z_l,\omega_m);\,
  \alpha \in\{{\rm x,y}\}.\\
 \label{e32}
 \end{eqnarray}
The relations~\refeq{e31} and \refeq{e32} assume that both ${\rm
x}$ and ${\rm y}$ polarization vectors $\ve{e}_{m,\alpha}$ of the
electric field [see Eq.~\refeq{e2}] preserve their orientation
after reflection at the boundary. This definition is equivalent to
that of the TE-polarized electric-field vector in the general
linear transmission/reflection scheme \cite{Yeh1988}. We note
that, in the analyzed 1D geometry, the TE- and TM-polarized waves
are physically equivalent.

To allow for further manipulations with the above derived
relations (and later the numerical treatment), we introduce
suitable orthonormal bases $f_{{\rm s},k}(\omega_{\rm s})$ and
$f_{{\rm i},k}(\omega_{\rm i})$, $ k=0,\ldots,\infty $, in the
signal and idler fields, respectively. This results in the
replacement of 'continuous indices' $ \omega_{\rm s} $ and $
\omega_{\rm i} $ by the discrete index $ k $. In these bases, new
signal [idler] field operators $\hat{A}_{{\rm
s}_a,\alpha,k}^{(l)}(z)$ [$\hat{A}_{{\rm
i}_b,\beta,k}^{(l)\dagger}(z)$] are defined as follows:
\begin{eqnarray} 
\label{e18}
 \hat{A}_{{\rm s}_{a},\alpha,k}^{(l)}(z) &=& \int_0^{\infty} d\omega_{\rm s}\,
  f_{{\rm s},k}^*(\omega_{\rm s})\,\hat{a}_{{\rm s}_a,\alpha}^{(l)}(z,\omega_{\rm s}).
\end{eqnarray}
The original operators are obtained by the inverse transformation
\begin{eqnarray} 
\label{e20}
 \hat{a}_{{\rm s}_a,\alpha}^{(l)}(z,\omega_{\rm s}) &=& \sum_{k=0}^{\infty} f_{{\rm s},k}(\omega_{\rm s})
  \hat{A}_{{\rm s}_{a},\alpha,k}^{(l)}(z).
\end{eqnarray}

Equations~\refeq{e31} and \refeq{e32} can be conveniently
rewritten into a matrix form. As the derivation procedure is
similar for both equations, we focus here only on the
transformation of Eq.~\refeq{e31} written for the signal field.
The solution for signal-field operator $\hat{a}_{{\rm
s}_a,\alpha}^{(l)}(z_l,\omega_{\rm s})$ given in Eq.~\refeq{e11}
is inserted into Eq.~\refeq{e31} first. Then, introducing
vectorial operators $\hat{\ve{A}}_{{\rm s}_{a},\alpha}^{(l)}$ and
$\hat{\ve{A}}_{{\rm i}_{b},\beta}^{(l)\dagger}$ with the elements
$[\hat{\ve{A}}_{{\rm s}_{a},\alpha}^{(l)}]_k = \hat{A}_{{\rm
s}_{a},\alpha,k}^{(l)}$ and $[\hat{\ve{A}}_{{\rm
i}_{b},\beta}^{(l)\dagger}]_k = \hat{A}_{{\rm
i}_{b},\beta,k}^{(l)\dagger}$ the relations~\refeq{e31} for the
continuity of electric-field amplitudes are expressed in the form:
\begin{align} 
\nonumber
 &\ve{I}_{E{\rm s},\alpha}^{(l-1)} \sum_{a={\rm F,B}}\hat{\ve{A}}_{{\rm s}_a,\alpha}^{(l-1)}(z_l) +
  \sum_{a,b={\rm F,B}}^{\beta={\rm x,y}}\ve{I}_{{\rm s},\alpha}^{(l-1)}\ve{J}_{E{\rm s},ab}^{\alpha\beta,(l-1)}(z_l)\\
\nonumber
 &\times \hat{\ve{A}}_{{\rm i}_b,\beta}^{(l-1) \dagger}(z_l) = \ve{I}_{E{\rm s},\alpha}^{(l)} \sum_{a={\rm F,B}}
  \hat{\ve{A}}_{{\rm s}_a,\alpha}^{(l)}(z_l)
  + \sum_{a,b={\rm F,B}}^{\beta={\rm x,y}}\ve{I}_{{\rm s},\alpha}^{(l)} \\
\label{e34} &\times\ve{J}_{E{\rm
 s},ab}^{\alpha\beta,(l)}(z_l)\hat{\ve{A}}_{{\rm i}_b,\beta}^{(l)
\dagger}(z_l).
\end{align}
The elements of matrices $\mathbf{I}^{(l)}_{E{\rm s},\alpha}$,
$\mathbf{I}^{(l)}_{{\rm s},\alpha}$ and $\mathbf{J}^{\alpha
\beta,(l)}_{E{\rm s},ab}$ found in Eq.~\refeq{e34} are defined as
\begin{align} 
 \bigl[\mathbf{I}^{(l)}_{{\rm s},\alpha}\bigr]_{kn} = \bigl[\mathbf{I}^{(l)}_{E{\rm s},\alpha}\bigr]_{kn}
  &\equiv \int_{0}^{\infty} d\omega_{\rm s} \frac{f_{{\rm s},k}^\ast(\omega_{\rm s})
  f_{{\rm s},n}(\omega_{\rm s})}{\sqrt{n^{(l)}_{{\rm s},\alpha}(\omega_{\rm s})}} , \\
 \bigl[\mathbf{J}_{E{\rm s},ab}^{\alpha\beta,(l)}\bigr]_{kn}(z) &\equiv
  \lambda_{E{\rm s},ab,kn}^{\alpha\beta,(l)}(z)
\label{e34b}
\end{align}
and the expansion coefficients $\lambda_{E{\rm
s},mn,ab}^{\alpha\beta,(l)}(z)$ are introduced according to the
relation
\begin{equation} 
 \Phi_{{\rm s},ab}^{\alpha \beta, (l)*}(z,\omega_{\rm s},\omega_{\rm i}) = \sum_{k,n=0}^{\infty}
  \lambda_{E{\rm s},ab,kn}^{\alpha \beta, (l)}(z) f_{{\rm s},k}(\omega_{\rm s}) f_{{\rm i},n}(\omega_{\rm i}).
 \label{e34bb}
\end{equation}
Equation~\refeq{e31} written for the idler field can be recast
into the form of Eq.~\refeq{e34} similarly. Equations~\refeq{e32}
written for the signal and idler fields can be rearranged into the
form of Eq.~\refeq{e34} as well. All four equations are then used
together to describe photon-pair generation.

Equation~\refeq{e34} can be divided into three independent
equations according to the physical origin of individual terms:
The first equation describes the linear field's transformation at
a boundary, the second equation governs photon pairs emitted in
the volumes of the $ (l-1) $-th and $ l $-th layers and the third
equation is appropriate for photon pairs born at the boundary
between the $ (l-1) $-th and $ l $-th layers in surface SPDC. We
note that the propagation index $a$ together with the layer number
($ l-1 $, $ l $) separate in Eq.~\refeq{e34} the terms describing
the fields impinging on the boundary (ingoing) from those leaving
the boundary (outgoing). In Eq.~\refeq{e34}, there occur the
free-field idler creation operators $\hat{\ve{A}}_{{\rm
i}_b,\beta}^{(l')\dagger} $ in the terms arising in the particular
solution. Their spatial evolution is described by the homogeneous
solution (without the nonlinear interaction) and so we denote them
by the upper index 0 ($ \hat{\ve{A}}_{{\rm
i}_b,\beta}^{(l'),0\dagger}$).

\begin{figure}   
 \centering
 \includegraphics[scale=1]{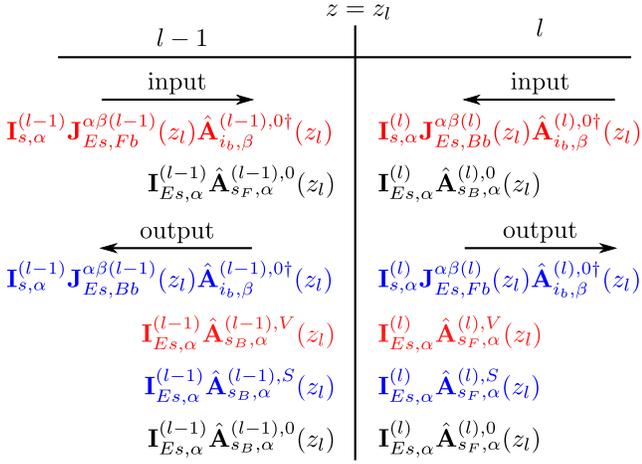}
 \caption{[Color online] Different terms occurring in Eq.~\refeq{e34} written for the
  boundary between $ (l-1) $-th and $ l $-th layers and the signal field.
  The upper index $ w $ in the operators
  $\hat{\ve{A}}_{m_a,\alpha}^{(l),w}$, $ m \in \{{\rm s,i}\} $, identifies volume-emitted
  photon pairs ($w={\rm V}$, red color), surface-emitted photon pairs ($w={\rm S}$, blue color) and linear
  propagation ($w=0$, black color).}
 \label{f-rozhrani}
\end{figure}

The operators ($\hat{\ve{A}}_{{\rm s}_{\rm F},\alpha}^{(l)}$ and
$\hat{\ve{A}}_{{\rm s}_{\rm B},\alpha}^{(l-1)}$) that describe in
Eq.~\refeq{e34} the fields propagating away from the boundary, can
be decomposed into three additive terms characterizing linear
transmission $ \hat{\ve{A}}_{{\rm s}_a,\alpha}^{(l'),0} $, photon
pairs generated in the volume $ \hat{\ve{A}}_{{\rm
s}_a,\alpha}^{(l'),{\rm V}} $ and photons pairs coming from the
boundary $ \hat{\ve{A}}_{{\rm s}_a,\alpha}^{(l'),{\rm S}} $:
\begin{eqnarray}   
\nonumber \hat{\ve{A}}_{{\rm s}_a,\alpha}^{(l')} &=&
\hat{\ve{A}}_{{\rm s}_a,\alpha}^{(l'),0} + \hat{\ve{A}}_{{\rm
s}_a,\alpha}^{(l'),{\rm V}} + \hat{\ve{A}}_{{\rm
s}_a,\alpha}^{(l'),{\rm S}};\\
&& \hspace{-3mm} (l'=l-1\wedge a={\rm B})\vee(l'=l\wedge a={\rm
F}). \label{e-rozdeleni-amplitud}
\end{eqnarray}
Inserting Eq.~\refeq{e-rozdeleni-amplitud} into Eq.~\refeq{e34},
we arrive at different terms that are identified with volume SPDC,
surface SPDC and linear transition at the boundary. Identification
and separation of different terms in Eq.~\refeq{e34} according to
the field's direction of propagation is shown in
Fig.~\ref{f-rozhrani}.

Volume SPDC as well as the linear propagation are described by one
input term and one output term for each field in each layer. On
the other hand, surface SPDC is described only by two output terms
for both the signal and idler fields at each boundary. These
fields arising in surface SPDC represent a nonlinear correction to
the usual Fresnel relations at the boundaries (see
\cite{Blombergen1962,Blombergen1969}) valid for linear materials.
The equations arising in the separation of different terms in
Eq.~\refeq{e34} are written in the form:
\begin{widetext}
\begin{eqnarray}   
\label{e-E-rozhrani-linearni} \ve{I}_{E{\rm s},\alpha}^{(l-1)}
\sum_{a={\rm F,B}} \hat{\ve{A}}_{{\rm s}_a,\alpha}^{(l-1),0}(z_l)
&=&
 \ve{I}_{E{\rm s},\alpha}^{(l)} \sum_{a={\rm F,B}} \hat{\ve{A}}_{{\rm s}_a,\alpha}^{(l),0}(z_l), \\
\label{e-E-rozhrani-emise-objem} \ve{I}_{E{\rm s},\alpha}^{(l-1)}
\hat{\ve{A}}_{{\rm s}_{\rm B},\alpha}^{(l-1),{\rm V}}(z_l) +
\ve{I}_{{\rm s},\alpha}^{(l-1)}\sum_{b={\rm F,B}}^{\beta={\rm
x,y}} \ve{J}_{E{\rm s},{\rm F}b}^{\alpha\beta,(l-1)}(z_l)
\hat{\ve{A}}_{{\rm i}_b,\beta}^{(l-1),0\dagger}(z_l) &=&
\ve{I}_{E{\rm s},\alpha}^{(l)}
 \hat{\ve{A}}_{{\rm s}_{\rm F},\alpha}^{(l),{\rm V}}(z_l) + \ve{I}_{{\rm s},\alpha}^{(l)}\sum_{b={\rm F,B}}^{\beta={\rm x,y}}\ve{J}_{E{\rm s},{\rm B}b}^{\alpha\beta,(l)}(z_l)
  \hat{\ve{A}}_{{\rm i}_b,\beta}^{(l),0\dagger}(z_l) ,\\
\label{e-E-rozhrani-emise-povrch} \ve{I}_{E{\rm s},\alpha}^{(l-1)}
 \hat{\ve{A}}_{{\rm s}_{\rm B},\alpha}^{(l-1),{\rm S}}(z_l) +
  \ve{I}_{{\rm s},\alpha}^{(l-1)}\sum_{b={\rm F,B}}^{\beta={\rm x,y}}\ve{J}_{E{\rm s},{\rm B}b}^{\alpha\beta,(l-1)}(z_l)
  \hat{\ve{A}}_{{\rm i}_b,\beta}^{(l-1),0 \dagger}(z_l) &=&
  \ve{I}_{E{\rm s},\alpha}^{(l)} \hat{\ve{A}}_{{\rm s}_{\rm F},\alpha}^{(l),{\rm S}}(z_l) +
  \ve{I}_{{\rm s},\alpha}^{(l)}\sum_{b={\rm F,B}}^{\beta={\rm x,y}}
  \ve{J}_{E{\rm s},{\rm F}b}^{\alpha\beta,(l)}(z_l)
  \hat{\ve{A}}_{{\rm i}_b,\beta}^{(l),0 \dagger}(z_l).
\end{eqnarray}
\end{widetext}

If we consider only the requirement of electric-field continuity
on the boundary and omit that for the magnetic field, no surface
SPDC would occur. This follows from the solution for signal-field
operators $\hat{a}_{{\rm s}_a,\alpha}(z,\omega_{\rm s})$ given in
Eq.~\refeq{e11}. The functions $\Phi_{{\rm s},{\rm B}b}^{\alpha
\beta, (l-1)}(z_{l},\omega_{\rm s},\omega_{\rm i})$ and
$\Phi_{{\rm s},{\rm F}b}^{\alpha \beta, (l)}(z_{l},\omega_{\rm
s},\omega_{\rm i})$ equal zero [see Eq.~\refeq{e13}] and so the
matrices $\ve{J}_{{\rm s},{\rm B}b}^{\alpha \beta,(l-1)}$ and
$\ve{J}_{{\rm s},{\rm F}b}^{\alpha \beta,(l)}$ equal zero.
Equations~\refeq{e-E-rozhrani-emise-povrch} thus separate from the
remaining two Eqs.~(\ref{e-E-rozhrani-linearni}) and
(\ref{e-E-rozhrani-emise-objem}) and the operators
$\hat{\ve{A}}_{{\rm s}_{\rm B},\alpha}^{(l-1),{\rm S}}$ and
$\hat{\ve{A}}_{{\rm s}_{\rm F},\alpha}^{(l),{\rm S}}$ describing
the surface emission could be set to zero. However, the continuity
of the magnetic field requires nonzero operators
$\hat{\ve{A}}_{{\rm s}_{\rm B},\alpha}^{(l-1),{\rm S}}$ and
$\hat{\ve{A}}_{{\rm s}_{\rm F},\alpha}^{(l),{\rm S}}$. These
operators then describe the surface emission of photon pairs at
the boundary.

The requirement of continuity for the magnetic field
$\hat{\ve{H}}_{\rm s}(z,t)$ across the boundary between the
$(l-1)$-th and $l$-th layers [see Eq.~\refeq{e32}] results in the
system of equations of the form written in
Eqs.~(\ref{e-E-rozhrani-linearni}---\ref{e-E-rozhrani-emise-povrch}).
These equations are formally derived from those written in
Eqs.~(\ref{e-E-rozhrani-linearni}---\ref{e-E-rozhrani-emise-povrch})
if we replace matrix $\ve{I}_{E{\rm s},\alpha}^{(l)} $ by matrix $
\ve{I}_{H{\rm s}_a,\alpha}^{(l)} $ and matrix $\ve{J}_{E{\rm
s},ab}^{\alpha\beta,(l)}(z) $ by matrix $ \ve{J}_{H{\rm
s},ab}^{\alpha\beta,(l)}(z) $. The matrices $ \ve{I}_{H{\rm
s}_a,\alpha}^{(l)} $ and $ \ve{J}_{H{\rm
s},ab}^{\alpha\beta,(l)}(z) $ are defined as
\begin{align} 
\label{e-K-matice-def}
 &\bigl[\ve{I}_{H{\rm s}_a,\alpha}^{(l)}\bigr]_{kn}
  \equiv i \int_0^{\infty}d\omega_{\rm s}\,\frac{k_{{\rm s}_a,\alpha}^{(l)}(\omega_{\rm s})
  f_{{\rm s},k}^\ast(\omega_{\rm s}) f_{{\rm s},n}(\omega_{\rm s}) }{\sqrt{n_{{\rm s},\alpha}^{(l)}(\omega_{\rm s})}} ,\\
\label{e-L-matice-def}
 &\bigl[\ve{J}_{H{\rm s},ab}^{\alpha\beta,(l)}
 \bigr]_{kn}(z) \equiv \lambda_{H{\rm s},ab,kn}^{\alpha\beta,(l)}(z)
\end{align}
and we assume the following decomposition
\begin{equation}  
\frac{\partial \Phi_{{\rm s},ab}^{\alpha \beta,(l)*}}{\partial
z}(z,\omega_{\rm s},\omega_{\rm i}) =
 \sum_{k,n=0}^{\infty} \lambda_{H{\rm s},ab,kn}^{\alpha\beta,(l)}(z) f_{{\rm s},k}(\omega_{\rm s}) f_{{\rm i},n}(\omega_{\rm i}).
\end{equation}

The boundary conditions for the free signal-field operators
$\hat{\ve{A}}_{{\rm s}_a,\alpha}^{(l'),0} ;\,a\in\{{\rm
F,B}\},\alpha\in\{{\rm x,y}\}$ arising from the continuity of the
electric- [Eq.~\refeq{e-E-rozhrani-linearni}] and magnetic-field
amplitudes and considered for both polarizations along the $ {\rm
x} $ and $ {\rm y} $ axes (the Fresnel relations) can be written
in the following compact form
\begin{equation}   
\label{e-E-rozhrani-linearni-final}
 \ve{L}_{\rm s}^{(l-1)} \hat{\ve{A}}_{\rm s}^{(l-1),0}(z_l) = \ve{L}_{\rm s}^{(l)}
 \hat{\ve{A}}_{\rm s}^{(l),0}(z_l)
\end{equation}
using the interface transition matrices $ \ve{L}_{\rm s}^{(l)} $
\cite{Yeh1988}:
\begin{equation}  
 \ve{L}_{\rm s}^{(l)} =
 \left[
 \begin{array}{cccc}
  \ve{I}_{E{\rm s,x}}^{(l)} & \ve{I}_{E{\rm s,x}}^{(l)} & 0 & 0 \\
  0 & 0 & \ve{I}_{E{\rm s,y}}^{(l)} & \ve{I}_{E{\rm s,y}}^{(l)} \\
  \ve{I}_{H{\rm s}_{\rm F},{\rm x}}^{(l)} & \ve{I}_{H{\rm s}_{\rm B},{\rm x}}^{(l)} & 0 & 0 \\
  0 & 0 & \ve{I}_{H{\rm s}_{\rm F},{\rm y}}^{(l)} & \ve{I}_{H{\rm s}_{\rm B},{\rm y}}^{(l)}
 \end{array}
 \right].
\end{equation}
In Eq.~\refeq{e-E-rozhrani-linearni-final} the signal-field
operators $\hat{\ve{A}}_{\rm s}^{(l),0}$ are defined as follows:
\begin{equation}  
 \hat{\ve{A}}_{\rm s}^{(l),0}(z) \equiv \left[
\begin{array}{c}
\hat{\ve{A}}_{{\rm s}_{\rm F},{\rm x}}^{(l),0}(z)\\
\hat{\ve{A}}_{{\rm s}_{\rm B},{\rm x}}^{(l),0}(z)\\
\hat{\ve{A}}_{{\rm s}_{\rm F},{\rm y}}^{(l),0}(z)\\
\hat{\ve{A}}_{{\rm s}_{\rm B},{\rm y}}^{(l),0}(z)
\end{array} \right] .
\end{equation}

The boundary conditions for the fields arising in volume SPDC and
considered for both electric-
[Eq.~\refeq{e-E-rozhrani-emise-objem}] and magnetic-field
operators can be expressed in the compact form of
Eq.~\refeq{e-E-rozhrani-linearni-final}:
\begin{eqnarray}  
\label{e-E-rozhrani-emise-objem-final}
 \ve{I}_{\rm s}(z_l)
  \hat{\ve{A}}_{\rm s}^{\rm V}(z_l) &=& -\ve{J}^{(l-1)}_{{\rm s}_{\rm F}}
  \hat{\ve{A}}_{\rm i}^{(l-1),0\dagger}(z_l)+ \ve{J}^{(l)}_{{\rm s}_{\rm B}}
  \hat{\ve{A}}_{\rm i}^{(l),0\dagger}(z_l) , \nonumber \\
 & &
\end{eqnarray}
where
\begin{equation}  
 \ve{I}_{\rm s}(z_l) \equiv \left[
\begin{array}{cccc}
 -\ve{I}_{E{\rm s,x}}^{(l)} & \ve{I}_{E{\rm s,x}}^{(l-1)} & 0 & 0 \\
 0 & 0 & -\ve{I}_{E{\rm s,y}}^{(l)} & \ve{I}_{E{\rm s,y}}^{(l-1)} \\
 -\ve{I}_{H{\rm s}_{\rm F},{\rm x}}^{(l)} & \ve{I}_{H{\rm s}_{\rm B},{\rm x}}^{(l-1)} & 0 & 0 \\
 0 & 0 & -\ve{I}_{H{\rm s}_{\rm F},{\rm y}}^{(l)} & \ve{I}_{H{\rm s}_{\rm B},{\rm y}}^{(l-1)}
\end{array}
\right]
\end{equation}
and
\begin{equation}  
 \hat{\ve{A}}_{\rm s}^{\rm V}(z_l) \equiv \left[
  \begin{array}{l}
  \hat{\ve{A}}_{{\rm s}_{\rm F},{\rm x}}^{(l),{\rm V}}(z_l)\\
  \hat{\ve{A}}_{{\rm s}_{\rm B},{\rm x}}^{(l-1),{\rm V}}(z_l)\\
  \hat{\ve{A}}_{{\rm s}_{\rm F},{\rm y}}^{(l),{\rm V}}(z_l)\\
  \hat{\ve{A}}_{{\rm s}_{\rm B},{\rm y}}^{(l-1),{\rm V}}(z_l)
  \end{array}
  \right] .
\end{equation}
Matrices $ \ve{J}^{(l)}_{{\rm s}_a} $, $ a \in \{{\rm F,B}\} $,
occurring in Eq.~\refeq{e-E-rozhrani-emise-objem-final} are given
as
\begin{eqnarray}  
 \ve{J}^{(l)}_{{\rm s}_a} &\equiv& \left[
  \begin{array}{cccc}
  \ve{I}_{\rm s,x}^{(l)} \ve{J}_{E{\rm s},a{\rm F}}^{{\rm x x},(l)}(z_l) &
  \ve{I}_{\rm s,x}^{(l)} \ve{J}_{E{\rm s},a{\rm B}}^{{\rm x x},(l)}(z_l) \\
  \ve{I}_{\rm s,y}^{(l)} \ve{J}_{E{\rm s},a{\rm F}}^{{\rm y x},(l)}(z_l) &
  \ve{I}_{\rm s,y}^{(l)} \ve{J}_{E{\rm s},a{\rm B}}^{{\rm y x},(l)}(z_l) \\
 \ve{I}_{\rm s,x}^{(l)} \ve{J}_{H{\rm s},a{\rm F}}^{{\rm x x},(l)}(z_l) &
 \ve{I}_{\rm s,x}^{(l)} \ve{J}_{H{\rm s},a{\rm B}}^{{\rm x x},(l)}(z_l) \\
 \ve{I}_{\rm s,y}^{(l)} \ve{J}_{H{\rm s},a{\rm F}}^{{\rm y x},(l)}(z_l) &
 \ve{I}_{\rm s,y}^{(l)} \ve{J}_{H{\rm s},a{\rm B}}^{{\rm y x},(l)}(z_l)
 \end{array} \right. \nonumber \\
  & & \hspace{5mm} \left. \begin{array}{cccc}
  \ve{I}_{\rm s,x}^{(l)} \ve{J}_{E{\rm s},a{\rm F}}^{{\rm x y},(l)}(z_l) &
  \ve{I}_{\rm s,x}^{(l)} \ve{J}_{E{\rm s},a{\rm B}}^{{\rm x y},(l)}(z_l) \\
  \ve{I}_{\rm s,y}^{(l)} \ve{J}_{E{\rm s},a{\rm F}}^{{\rm y y},(l)}(z_l) &
  \ve{I}_{\rm s,y}^{(l)} \ve{J}_{E{\rm s},a{\rm B}}^{{\rm y y},(l)}(z_l) \\
 \ve{I}_{\rm s,x}^{(l)} \ve{J}_{H{\rm s},a{\rm F}}^{{\rm x y},(l)}(z_l) &
 \ve{I}_{\rm s,x}^{(l)} \ve{J}_{H{\rm s},a{\rm B}}^{{\rm x y},(l)}(z_l) \\
 \ve{I}_{\rm s,y}^{(l)} \ve{J}_{H{\rm s},a{\rm F}}^{{\rm y y},(l)}(z_l) &
 \ve{I}_{\rm s,y}^{(l)} \ve{J}_{H{\rm s},a{\rm B}}^{{\rm yy},(l)}(z_l)
 \end{array} \right].
\end{eqnarray}

Similarly, the boundary conditions for surface SPDC including both
electric- [Eq.~\refeq{e-E-rozhrani-emise-povrch}] and
magnetic-field operators and their polarizations are obtained in
the following compact form:
\begin{equation}   
\label{e-rozhrani-emise-povrch-final}
 \ve{I}_{\rm s}(z_l) \hat{\ve{A}}_{\rm s}^{\rm S}(z_l) = -\ve{J}^{(l-1)}_{{\rm s}_{\rm B}}
  \hat{\ve{A}}_{\rm i}^{(l-1),0\dagger}(z_l)+ \ve{J}^{(l)}_{{\rm s}_{\rm F}} \hat{\ve{A}}_{\rm i}^{(l),0\dagger}(z_l).
\end{equation}

Equations~\refeq{e-E-rozhrani-linearni-final},
\refeq{e-E-rozhrani-emise-objem-final} and
\refeq{e-rozhrani-emise-povrch-final} characterize the behavior of
the overall signal field at the boundaries. The idler field
behaves at the boundaries in the same way and the corresponding
equations characterizing its behavior are derived in the same form
as those for the signal field. We remind that they are obtained
from Eqs.~\refeq{e-E-rozhrani-linearni-final},
\refeq{e-E-rozhrani-emise-objem-final} and
\refeq{e-rozhrani-emise-povrch-final} by formal substitution $ s
\leftrightarrow i $. The equations for the signal and idler fields
are mutually coupled and so we have to solve them together. For
this reason, we first express them in the following 'super-vector'
and 'super-matrix' notation:
\begin{align}  
\label{e-linearni-prechod-rozhrani-final-2}
&\mathcal{L}^{(l-1)} \hat{\mathcal{A}}^{(l-1),0}(z_l) = \mathcal{L}^{(l)}\hat{\mathcal{A}}^{(l),0}(z_l),\\
\label{e-emise-objem-final-2} &\mathcal{I}(z_l)
\hat{\mathcal{A}}^{\rm V}(z_l) = -\mathcal{J}_{\rm F}^{(l-1)}
\hat{\mathcal{A}}^{(l-1),0}(z_l) +
\mathcal{J}_{\rm B}^{(l)} \hat{\mathcal{A}}^{(l),0}(z_l),\\
\label{e-rozhrani-emise-final-2} &\mathcal{I}(z_l)
\hat{\mathcal{A}}^{\rm S}(z_l) = - \mathcal{J}_{\rm B}^{(l-1)}
\hat{\mathcal{A}}^{(l-1),0}(z_l) + \mathcal{J}_{\rm F}^{(l)}
\hat{\mathcal{A}}^{(l),0}(z_l)
\end{align}
that uses the 'super-matrices' $\mathcal{L}^{(l)}$, $\mathcal{I}$
and $\mathcal{J}_a$,
\begin{eqnarray}  
\label{e-D-matice-definice}
 \mathcal{L}^{(l)} &\equiv& \mathrm{diag} \left[\ve{L}_{\rm s}^{(l)},\ve{L}_{\rm i}^{(l)} \right],\\
 \label{e-I-matice-definice}
 \mathcal{I}(z_l)  &\equiv& \mathrm{diag}\Bigl[\ve{I}_{\rm s}(z_l),\ve{I}_{\rm i}(z_l)\Bigr],\\
  \label{e-Ja-matice-definice}
 \mathcal{J}_a^{(l)} &\equiv& \mathrm{adiag}\left[\ve{J}_{{\rm s}_a}^{(l)},\ve{J}_{{\rm i}_a}^{(l)} \right];\,a\in\{{\rm F,B}\}.
\end{eqnarray}
Symbol $ {\rm diag} $ ($ {\rm adiag} $) stands for a diagonal
(anti-diagonal) matrix. 'Super-vector' operators
$\hat{\mathcal{A}}^{(l),0}(z)$, $\hat{\mathcal{A}}^{\rm V}(z_l) $
and $\hat{\mathcal{A}}^{\rm S}(z_l) $ introduced in
Eqs.~(\ref{e-D-matice-definice}---\ref{e-Ja-matice-definice}) are
defined as
\begin{align}  
\label{e-definice-D0-celkovy-vektor}
\hat{\mathcal{A}}^{(l),0}(z)&= \left[
\begin{array}{c}
 \hat{\ve{A}}^{(l)}_{\rm s}(z) \\
 \hat{\ve{A}}^{(l)\dagger}_{\rm i}(z)
\end{array}
\right],\\
\hat{\mathcal{A}}^{w}(z_l)&= \left[
\begin{array}{c}
 \hat{\ve{A}}^{w}_{\rm s}(z_l) \\
 \hat{\ve{A}}^{w \dagger}_{\rm i}(z_l)
\end{array}
\right];\hspace{5mm} w \in\{{\rm S,V}\}.
\end{align}

The free-field operators $\hat{\mathcal{A}}^{(l),0} $, $ l \in
\{1,\ldots,N\}$, represent the source of photon-pairs emitted
either in the volume of layers [Eq.~\refeq{e-emise-objem-final-2}]
or at the boundaries between the layers
[Eq.~\refeq{e-rozhrani-emise-final-2}]. To quantify both volume
and surface SPDC, we need to express these operators in terms of
those impinging on the crystal. We first derive the relations
between operators $\hat{\mathcal{A}}^{(l),0}$ at the left and
right boundaries of an $ l $-th homogeneous layer. Using
Eqs.~\refeq{e14}, \refeq{e18} and \refeq{e20} we arrive at the
formula
\begin{equation} 
\label{e-pernosova-matice-skrze-vrstvu} \hat{\ve{A}}_{\rm
s}^{(l),0}(z_{l+1}) = \ve{P}_{\rm s}^{(l)} \hat{\ve{A}}_{\rm
s}^{(l),0}(z_l)
\end{equation}
in which the matrix $\ve{P}_{\rm s}^{(l)}$ is given as
\begin{eqnarray}  
\label{e-P-matice-definice}
 \ve{P}_{\rm s}^{(l)} \equiv
 \mathrm{diag}\left[\ve{P}_{{\rm s}_{\rm F},{\rm x}}^{(l)},\ve{P}_{{\rm s}_{\rm B},{\rm x}}^{(l)},
  \ve{P}_{{\rm s}_{\rm F},{\rm y}}^{(l)},\ve{P}_{{\rm s}_{\rm B},{\rm y}}^{(l)}\right],
\end{eqnarray}
and
\begin{eqnarray}  
 \bigl[\ve{P}_{{\rm s}_a,\alpha}^{(l)}\bigr]_{kn} = \int_0^{\infty} d\omega_{\rm s}\,\exp[ik_{{\rm s}_a,\alpha}^{(l)}(\omega_{\rm s})L_l]
  f_{{\rm s},k}^*(\omega_{\rm s}) f_{{\rm s},n}(\omega_{\rm s}). \nonumber \\
\end{eqnarray}
For 'super-vector' $\hat{\mathcal{A}}^{(l),0}$ containing both
signal- and idler-field operators,
Eq.~\refeq{e-pernosova-matice-skrze-vrstvu} attains the form
\begin{equation}  
 \hat{\mathcal{A}}^{(l),0}(z_{l+1}) = \mathcal{P}^{(l)} \hat{\mathcal{A}}^{(l),0}(z_l)
\end{equation}
and
\begin{equation}  
 \mathcal{P}^{(l)} =
 \left[
 \begin{array}{cc}
 \ve{P}_{\rm s}^{(l)} & 0 \\
 0 & \ve{P}_{\rm i}^{(l)}
 \end{array}
 \right] .
\end{equation}

The operators $\hat{\mathcal{A}}^{(0),0}(z_1)$ at the left-hand
side of the structure and operators
$\hat{\mathcal{A}}^{(N+1),0}(z_{N+1})$ at the right-hand side of
the structure are related by the following equation
\begin{equation} 
\label{e56}
 \hat{\mathcal{A}}^{(N+1),0}(z_{N+1})=\mathcal{T}^{(N+1,0)} \hat{\mathcal{A}}^{(0),0}(z_{1})
\end{equation}
where
\begin{eqnarray}  
 \nonumber
  \mathcal{T}^{(n,m)} &\equiv& \mathcal{D}^{(n)-1}
  \left(\prod_{l={m+1}^n-1} \mathcal{D}^{(l)} \mathcal{P}^{(l)} \mathcal{D}^{(l)-1}\right) \mathcal{D}^{(m)};\\
 \label{e57}
 & & \hspace{5mm}  m,n\in\{0,\ldots,N+1\},n>m.
\end{eqnarray}
In Eq.~\refeq{e57}, terms in the product are multiplied from the
right to the left as index $l$ increases. If $n=m+1$, the product
in Eq.~\refeq{e57} is set to unity by definition. The general
transfer matrix $\mathcal{T}^{(n,m)}$ defined in Eq.~\refeq{e57}
transfers the operators from the right boundary of layer $m$
($z=z_{m}$) to the left boundary of layer $n$ ($z=z_{n-1}$) (see
Fig.~\ref{f1}). The matrix $\mathcal{T}^{(N+1,0)}$ then describes
the propagation through the whole layered structure.

The operators $\hat{\ve{A}}_{a,\alpha}^{(N+1),0}(z_{N+1})$ and
$\hat{\ve{A}}_{a,\alpha}^{(0),0}(z_{1})$ embedded in the
'super-vectors' $\hat{\mathcal{A}}^{(N+1),0}(z_{N+1})$ and
$\hat{\mathcal{A}}^{(0),0}(z_{1})$, respectively, have to be
rearranged to express the field operators leaving the structure
[$\hat{\ve{A}}_{{\rm F},\alpha}^{(N+1),0}(z_{N+1})$ and
$\hat{\ve{A}}_{{\rm B},\alpha}^{(0),0}(z_{1})$] in terms of the
operators entering the structure [$\hat{\ve{A}}_{{\rm
F},\alpha}^{(0),0}(z_{1})$ and $\hat{\ve{A}}_{{\rm
B},\alpha}^{(N+1),0}(z_{N+1})$]. As the relations among the
considered field operators are linear, the needed formulas are
easily found. To express them, we introduce the notation in which
the signal- and idler-field operators are suitably rearranged:
\begin{equation}  
  \hat{\mathcal{A}}^{{\rm (out)},w} = \left[
  \begin{array}{l}
  \hat{\ve{A}}_{\rm F,x}^{(N+1),w}\\
  \hat{\ve{A}}_{\rm B,x}^{(0),w}\\
  \hat{\ve{A}}_{\rm F,y}^{(N+1),w}\\
  \hat{\ve{A}}_{\rm B,y}^{(0),w}
  \end{array}
  \right],\, \hat{\mathcal{A}}^{{\rm (in)},w} = \left[
 \begin{array}{l}
  \hat{\ve{A}}_{\rm F,x}^{(0),w}\\
  \hat{\ve{A}}_{\rm B,x}^{(N+1),w}\\
  \hat{\ve{A}}_{\rm F,y}^{(0),w}\\
  \hat{\ve{A}}_{\rm B,y}^{(N+1),w}
 \end{array}
\right]
\end{equation}
and
\begin{eqnarray} 
 \hat{\ve{A}}_{a,\alpha}^{(l),w}(z) &\equiv& \left[
 \begin{array}{c}
  \hat{\ve{A}}_{{\rm s}_a,\alpha}^{(l),w} (z)\\
  \nonumber
  \hat{\ve{A}}_{{\rm i}_a,\alpha}^{(l),w\dagger} (z)
 \end{array}
  \right], \hspace{5mm} w \in \{0,{\rm S,V}\}, \nonumber \\
 & & \hspace{5mm} \,a\in\{{\rm F,B}\},\alpha \in
 \{{\rm x,y}\}.
\end{eqnarray}
In this notation, the input-output formulas for the fields'
operators are written as
\begin{equation} 
 \label{e58}
 \hat{\mathcal{A}}^{\rm (out),0}=\mathcal{F}\, \hat{\mathcal{A}}^{\rm
 (in),0}.
\end{equation}
Detailed calculations reveal the matrix $ \mathcal{F} $ in the
form:
\begin{eqnarray}   
 \mathcal{F} &\equiv& \mathcal{U}^{-1}\mathcal{V},
\label{e67} \\
 & & \mathcal{U}\equiv \left[
 \begin{array}{rrrr}
  1 & -[\mathcal{T}]_{\rm Fx,Bx} & 0 & -[\mathcal{T}]_{\rm Fx,By}\\
  0 & -[\mathcal{T}]_{\rm Bx,Bx} & 0 & -[\mathcal{T}]_{\rm Bx,By}\\
  0 & -[\mathcal{T}]_{\rm Fy,Bx} & 1 & -[\mathcal{T}]_{\rm Fy,By}\\
  0 & -[\mathcal{T}]_{\rm By,Bx} & 0 & -[\mathcal{T}]_{\rm By,By}
 \end{array} \right], \nonumber \\
 & & \mathcal{V}\equiv \left[
 \begin{array}{rrrr}
  [\mathcal{T}]_{\rm Fx,Fx} & 0 & [\mathcal{T}]_{\rm Fx,Fy} & 0\\
  \left[\mathcal{T}\right]_{\rm Bx,Fx} & - 1 & [\mathcal{T}]_{\rm Bx,Fy} & 0\\
  \left[\mathcal{T}\right]_{\rm Fy,Fx} & 0 & [\mathcal{T}]_{\rm Fy,Fy} & 0\\
  \left[\mathcal{T}\right]_{\rm By,Fx} & 0 & [\mathcal{T}]_{\rm By,Fy} & - 1
 \end{array} \right].  \nonumber
\end{eqnarray}
We remind that the operators $\hat{\ve{A}}_{a,\alpha}^{(N+1),0}$
at the right-hand side of the structure are evaluated at position
$z_{N+1}$, whereas the operators $\hat{\ve{A}}_{a,\alpha}^{(0),0}$
at the left-hand side of the structure are determined at position
$z_{1}$. In the definitions of matrices $ \mathcal{U} $ and $
\mathcal{V} $ in Eq.~\refeq{e67}, symbol $ 1 $ means the diagonal
unity matrix of appropriate dimensions.

Utilizing transformations \refeq{e56} and \refeq{e58} the
operators $\mathcal{\hat{A}}^{(l),0}(z_{l})$ in layer $l$ are
expressed in terms of the input operators as
\begin{equation}  
 \label{e-transformace-vrstva-vstup}
  \hat{\mathcal{A}}^{(l),0}(z_{l}) =
  \mathcal{T}^{(l,0)}\mathcal{W}\, \hat{\mathcal{A}}^{\rm (in),0}
\end{equation}
using matrix $ \mathcal{W} $ defined as
\begin{equation}  
 \mathcal{W} \equiv \left[
  \begin{array}{cccc}
   1 & 0 & 0 & 0\\
   \left[\mathcal{F}\right]{}_{\rm Bx,Fx} & \left[\mathcal{F}\right]{}_{\rm Bx,Bx} &
     \left[\mathcal{F}\right]{}_{\rm Bx,Fy} & \left[\mathcal{F}\right]{}_{\rm Bx,By}\\
   0 & 0 & 1 & 0 \\
   \left[\mathcal{F}\right]_{\rm By,Fx} & \left[\mathcal{F}\right]_{\rm By,Bx} & \left[\mathcal{F}\right]_{\rm By,Fy} &
   \left[\mathcal{F}\right]_{\rm By,By}
  \end{array} \right] .
\end{equation}
Exploiting Eqs.~\refeq{e-pernosova-matice-skrze-vrstvu} and
\refeq{e-transformace-vrstva-vstup} the operators
$\mathcal{\hat{A}}^{(l),0}(z_{l+1})$ are expressed via the input
operators along the relation:
\begin{equation}  
\label{e-transformace-konec-vrstvy-vstup}
 \mathcal{\hat{A}}^{(l),0}(z_{l+1}) = \mathcal{P}^{(l)}
 \mathcal{T}^{(l,0)}\mathcal{W}\,\hat{\mathcal{A}}^{\rm (in),0}.
\end{equation}

The operators $\hat{\mathcal{A}}^{\rm V}(z_l)$ and
$\hat{\mathcal{A}}^{\rm S}(z_l)$ determined in
Eqs.~\refeq{e-emise-objem-final-2} and
\refeq{e-rozhrani-emise-final-2}, respectively, describe photons
born in volume and surface SPDC. Such photons, after being emitted
in a given layer or at a given boundary, propagate as free fields
towards the output planes of the structure. This propagation obeys
the following linear relations:
\begin{equation}  
\label{e-sireni-vystup-vystup}
 \hat{\mathcal{A}}^w(z_l) = \mathcal{X}(z_l) \mathcal{Y}\, \hat{\mathcal{A}}^{{\rm (out)},w};\, w \in
  \{{\rm S,V}\}.
\end{equation}
In Eq.~\refeq{e-sireni-vystup-vystup}, matrices $ \mathcal{X}(z_l)
$ and $ \mathcal{Y} $ are defined along the relations:
\begin{eqnarray}  
 \label{e-X-matice-definice}
 \mathcal{X}(z_l) &\equiv&
 \left[
 \begin{array}{cc}
  \left[ \mathcal{T}^{(l,0)}\right]_{\rm Fx,Fx} & \left[ \mathcal{T}^{(l,0)}\right]_{\rm Fx,Bx} \\
  \left[ \tilde{\mathcal{T}}^{(l,0)}\right]_{\rm Bx,Fx} & \left[ \tilde{\mathcal{T}}^{(l,0)}\right]_{\rm Bx,Bx}\\
  \left[ \mathcal{T}^{(l,0)}\right]_{\rm Fy,Fx} & \left[ \mathcal{T}^{(l,0)}\right]_{\rm Fy,Bx} \\
  \left[ \tilde{\mathcal{T}}^{(l,0)}\right]_{\rm By,Fx} & \left[\tilde{\mathcal{T}}^{(l,0)}\right]_{\rm By,Bx}
 \end{array}  \right. \nonumber \\
  & & \hspace{5mm} \left.
 \begin{array}{cc}
  \left[ \mathcal{T}^{(l,0)}\right]_{\rm Fx,Fy} & \left[ \mathcal{T}^{(l,0)}\right]_{\rm Fx,By} \\
  \left[ \tilde{\mathcal{T}}^{(l,0)}\right]_{\rm Bx,Fy} & \left[ \tilde{\mathcal{T}}^{(l,0)}\right]_{\rm Bx,By}\\
  \left[ \mathcal{T}^{(l,0)}\right]_{\rm Fy,Fy} & \left[ \mathcal{T}^{(l,0)}\right]_{\rm Fy,By} \\
  \left[ \tilde{\mathcal{T}}^{(l,0)}\right]_{\rm By,Fy} &\left[ \tilde{\mathcal{T}}^{(l,0)}\right]_{\rm By,By}
 \end{array}  \right] , \\
 \label{e-Y-matice-definice}
 \mathcal{Y} &\equiv&
 \left[
 \begin{array}{cccc}
  \left[\mathcal{Z}\right]_{\rm Fx,Fx} & \left[\mathcal{Z}\right]_{\rm Fx,Bx} & \left[\mathcal{Z}\right]_{\rm Fx,Fy} & \left[\mathcal{Z}\right]_{\rm Fx,By} \\
  0 & 1 & 0 & 0 \\
  \left[\mathcal{Z}\right]_{\rm Fy,Fx} & \left[\mathcal{Z}\right]_{\rm Fy,Bx} & \left[\mathcal{Z}\right]_{\rm Fy,Fy} & \left[\mathcal{Z}\right]_{\rm Fy,By} \\
  0 & 0 & 0 & 1
  \end{array}
  \right] \nonumber .\\
 & &
\end{eqnarray}
In Eq.~\refeq{e-X-matice-definice}, $\tilde{\mathcal{T}}^{(l,0)} =
\mathcal{P}^{(l)} \mathcal{T}^{(l,0)}$. The matrix $\mathcal{Z}$
occurring in Eq.~\refeq{e-Y-matice-definice} stands for the
inverse matrix to $ \mathcal{F}$ defined in Eq.~\refeq{e67}.

Now we return back to the central
equations~\refeq{e-emise-objem-final-2} and
\refeq{e-rozhrani-emise-final-2} that describe the emission of
photon pairs around the boundary surrounded by the $ (l-1) $-th
and $ l $-th layers. Whereas Eq.~\refeq{e-emise-objem-final-2}
describes photons emitted in volume SPDC and propagating forward
in the $ (l-1) $-th layer and backward in the $ l $-th layer,
Eq.~\refeq{e-rozhrani-emise-final-2} characterizes photon pairs
coming from surface SPDC occurring at the boundary between the two
layers. The 'local' operators occurring in these equations have to
be replaced by those describing the fields outside the structure
and mutually related by free-field propagation. The operators of
the fields impinging on the boundary from the left- as well as
right-hand side are replaced by those entering the structure
applying Eq.~\refeq{e-transformace-vrstva-vstup}. On the other
hand, the operators characterizing the fields propagating out of
the boundary are substituted by those appropriate for the fields
leaving the whole structure with the help of
Eq.~\refeq{e-sireni-vystup-vystup}. This results in the relation
between the input and output operators of the fields describing
one photon pair born around the boundary of the $ (l-1) $-th and $
l $-th layers:
\begin{equation} 
 \hat{\mathcal{A}}^{{\rm (out)},w}_{(l)} = \mathcal{S}^{(l,l-1),w} \hat{\mathcal{A}}^{\rm (in),0};
 \hspace{3mm} w\in\{{\rm S,V}\}
\end{equation}
and
\begin{eqnarray} 
 \mathcal{S}^{(l,l-1),{\rm V}} &=& \left[\mathcal{I}(z_l) \mathcal{X}(z_l) \mathcal{Y} \right]^{-1}
  \nonumber \\
  & & \left(-\mathcal{J}_{\rm F}^{(l-1)} \tilde{\mathcal{T}}^{(l-1,0)} + \mathcal{J}_{\rm B}^{(l)}
   \mathcal{T}^{(l,0)}\right) \mathcal{W}, \nonumber \\
 \mathcal{S}^{(l,l-1),{\rm S}} &=& \left[\mathcal{I}(z_l) \mathcal{X}(z_l) \mathcal{Y} \right]^{-1}
  \nonumber \\
  & & \left(-\mathcal{J}_{\rm B}^{(l-1)} \tilde{\mathcal{T}}^{(l-1,0)} + \mathcal{J}_{\rm F}^{(l)}
   \mathcal{T}^{(l,0)}\right) \mathcal{W}.
\end{eqnarray}
The overall operators of the fields at the output of the structure
are given by coherent superposition of the contributions from all
layers and their boundaries:
\begin{eqnarray}  
\label{e-nelinearni-transformace-vstup-vystup}
 \hat{\mathcal{A}}^{{\rm (out)},w} &=& \mathcal{G}^w\,\hat{\mathcal{A}}^{\rm (in),0},
  \nonumber \\
  \mathcal{G}^w &=& \sum_{l=1}^{N+1}\mathcal{S}^{(l,l-1),w};\, w\in\{{\rm S,V}\}.
\end{eqnarray}

Knowing relation \refeq{e-nelinearni-transformace-vstup-vystup}
between the operators $ \hat{\mathcal{A}}^{{\rm (out)},w} $, $ w
\in \{{\rm S,V}\} $, and $ \hat{\mathcal{A}}^{\rm (in),0} $, the
application of transformations given in Eqs.~\refeq{e18} and
\refeq{e20} provides the formula expressing the output
signal-field operators $\hat{a}_{{\rm s}_a,\alpha}^{\rm (out)}$ in
terms of the input idler-field operators $\hat{\bar{a}}_{{\rm
i}_b,\beta}^{{\rm (in)}\dagger}$:
\begin{eqnarray}   
 \nonumber
 \hat{a}_{{\rm s}_a,\alpha}^{{\rm (out)},w}(\omega_{\rm s}) &=& \ve{f}_{\rm s}^{\rm T}(\omega_{\rm s})
  \sum_{b={\rm F,B}}^{\beta={\rm x,y}} \bigl[\mathcal{G}^w \bigr]_{{\rm s}_a\alpha,{\rm i}_b\beta}
   \int_0^{\infty} d\omega_{\rm i} \ve{f}_{\rm i}(\omega_{\rm i})\\
 \nonumber
 & & \times \hat{\bar{a}}_{{\rm i}_b,\beta}^{{\rm (in)},0\dagger}(\omega_{\rm i}); \hspace{3mm} w \in \{{\rm S,V}\}, \\
 \label{e-d-operator-vstup-vystup-nelinearni}
 & & \hspace{0mm} \,a\in\{{\rm F,B}\}, \alpha\in\{{\rm x,y}\}.
\end{eqnarray}
The output signal-field operators $\hat{a}_{{\rm s}_{\rm
F},\alpha}^{{\rm (out)},w} \equiv \hat{a}_{{\rm s}_{\rm
F},\alpha}^{(N+1),w}(z_{N+1},\omega_{\rm s}) $ [$\hat{a}_{{\rm
s}_{\rm B},\alpha}^{{\rm (out)},w} \equiv \hat{a}_{{\rm s}_{\rm
B},\alpha}^{(0),w}(z_1,\omega_{\rm s})$] are located at the
position $z=z_{N+1}$ [$z=z_1$]. On the other hand, the input
idler-field operators $\hat{\bar{a}}_{{\rm i}_{\rm F},\beta}^{\rm
(in),0} \equiv \hat{\bar{a}}_{{\rm i}_{\rm F},\beta}^{\rm
(0)}(z_{1},\omega_{\rm i})$ [$\hat{\bar{a}}_{{\rm i}_{\rm
B},\beta}^{\rm (in),0} \equiv \hat{\bar{a}}_{{\rm i}_{\rm
B},\beta}^{\rm (N+1)}(z_{N+1},\omega_{\rm i})$] characterize the
field at position $ z_1 $ [$ z_{N+1} $]. In
Eq.~\refeq{e-d-operator-vstup-vystup-nelinearni}, vectors
$[\ve{f}_m] (\omega_m) $ containing elements $ f_{m,k}(\omega_m)$
for $ m \in \{{\rm s,i}\} $ have been introduced, symbol
$\bigl[\mathcal{G}^w \bigr]_{{\rm s}_a\alpha,{\rm i}_b\beta}$
denotes a sub-matrix of matrix $\mathcal{G}^w$ identified by its
indices and symbol $ {\rm T} $ stands for transposition.

In parallel with Eq.~\refeq{e-d-operator-vstup-vystup-nelinearni},
the formula for linear propagation of the signal-field operators
$\hat{\bar{a}}_{{\rm s}_a,\alpha}^0$ from the input of the layered
structure to its output including scattering of the field at the
boundaries is determined from Eq.~\refeq{e58}:
 \begin{eqnarray}  
 \nonumber
 \hat{\bar{a}}_{{\rm s}_a,\alpha}^{\rm (out),0}(\omega_{\rm s}) &\equiv& \ve{f}_{\rm s}^{\rm T}(\omega_{\rm s})
  \sum_{b={\rm F,B}}^{\beta={\rm x,y}} \bigl[\mathcal{F} \bigr]_{{\rm s}_a\alpha,{\rm s}_b\beta}
  \int_0^{\infty} d\omega_{\rm s}' \ve{f}_{\rm s}^*(\omega_{\rm s}')\\
\label{e-d-operator-vstup-vystup-linearni}
 && \hspace{-5mm} \times \hat{\bar{a}}_{{\rm s}_b,\beta}^{\rm (in),0}(\omega_{\rm s}');
  \hspace{3mm}a\in\{{\rm F,B}\}, \alpha\in\{{\rm x,y}\}.
\end{eqnarray}

\section{Experimental characteristics of photon pairs}

The emitted photon pairs are characterized by the joint
signal-idler photon-number density $n_{ab}^{\alpha \beta}$ defined
along the relation:
\begin{eqnarray}  
 \nonumber
  n_{ab}^{\alpha \beta}(\omega_{\rm s},\omega_{\rm i}) &=& \langle\mathrm{vac}\vert
  \hat{a}_{{\rm s}_a,\alpha}^{{\rm (out)} \dagger}(\omega_{\rm s})
  \hat{a}_{{\rm s}_a,\alpha}^{\rm (out)} (\omega_{\rm s}) \\
  \label{e-korelacni-funkce-gamma}
  && \times \hat{a}_{{\rm i}_b,\beta}^{{\rm (out)} \dagger} (\omega_{\rm i})
   \hat{a}_{{\rm i}_b,\beta}^{\rm (out)} (\omega_{\rm i}) \vert \mathrm{vac}
   \rangle.
\end{eqnarray}
The photon-number density $ n_{ab}^{\alpha \beta}(\omega_{\rm
s},\omega_{\rm i}) $ gives the density of photon pairs with a
signal photon at frequency $ \omega_{\rm s} $ propagating in
direction $ a $ with polarization $ \alpha $ and its idler twin at
frequency $ \omega_{\rm i} $ propagating in direction $ b $ with
polarization $ \beta $. Assuming vacuum around the structure and
using Eqs.~\refeq{e-d-operator-vstup-vystup-nelinearni} and
\refeq{e-d-operator-vstup-vystup-linearni}, we arrive at the
formula:
\begin{eqnarray}  
 n_{ab}^{\alpha \beta}(\omega_{\rm s},\omega_{\rm i}) &=& \sum_{w,w'={\rm S,V}}\sum_{g={\rm F,B}}^{\gamma={\rm x,y}}
  \ve{f}_{\rm i}^{*,{\rm T}}(\omega_{\rm i}) \bigl[\mathcal{F}\bigr]_{{\rm i}_b\beta,{\rm i}_g\gamma} \nonumber  \\
 & & \hspace{-10mm} \times \bigl[\mathcal{G}^w\bigr]_{{\rm s}_a\alpha,{\rm i}_g\gamma}^{\rm T} \ve{f}_{\rm s}^*(\omega_{\rm s})
  \sum_{d={\rm F,B}}^{\delta={\rm x,y}}
  \ve{f}_{\rm s}^{\rm T}(\omega_{\rm s}) \bigl[\mathcal{F}\bigr]_{{\rm s}_a\alpha,{\rm s}_d\delta} \nonumber\\
 & & \hspace{-10mm} \times \bigl[\mathcal{G}^{w'}\bigr]_{{\rm i}_b\beta,{\rm s}_d\delta}^{*,\rm T} \ve{f}_{\rm i}(\omega_{\rm i}) + {\rm h.c.}  \nonumber \\
 &\equiv& n_{ab}^{\alpha \beta,{\rm V}}(\omega_{\rm s},\omega_{\rm i}) + n_{ab}^{\alpha
  \beta,{\rm S}}(\omega_{\rm s},\omega_{\rm i}) \nonumber \\
 & & + n_{ab}^{\alpha \beta,{\rm I}}(\omega_{\rm s},\omega_{\rm i})  \nonumber \\
 &\equiv&  n_{ab}^{\alpha \beta,{\rm SV}}.
\label{e-korelacni-funkce-gamma-final}
\end{eqnarray}
According to Eq.~\refeq{e-korelacni-funkce-gamma-final}, the joint
photon-number density $n_{ab}^{\alpha \beta}$ is decomposed into
three contributions: The first contribution
$n_{ab}^{\alpha\beta,{\rm V}}$ originates in volume SPDC [$
w=w'={\rm V}$ in the sum at the first line of
Eq.~\refeq{e-korelacni-funkce-gamma-final}], the second
contribution $n_{ab}^{\alpha \beta,{\rm S}}$ arises in surface
SPDC ($ w=w'={\rm S}$) and the last contribution $n_{ab}^{\alpha
\beta,{\rm I}}$ occurs due to interference between the volume and
surface contributions. However, these contributions cannot be
mutually separated in the considered 1D model in the experiment.
That is why the three contributions summed together give the
overall joint photon-number density $n_{ab}^{\alpha \beta,{\rm
SV}}$.

The signal photon-number density $n_{s,ab}^{\alpha \beta,w} $
defined for $ w\in\{{\rm S,V,SV}\}$ is derived along the relation
\begin{equation}  
 n_{{\rm s},ab}^{\alpha \beta,w}(\omega_{\rm s}) = \int_{0}^{\infty}d\omega_{\rm i}\,
  n_{ab}^{\alpha \beta,w}(\omega_{\rm s},\omega_{\rm i}).
\end{equation}
Similarly, the number $N_{ab}^{\alpha \beta,w}$ of emitted photon
pairs is given by the formula
\begin{equation} 
\label{e69}
 N_{ab}^{\alpha\beta,w} = \int_0^{\infty} d\omega_{\rm s}\, n_{{\rm s},ab}^{\alpha \beta,w}(\omega_{\rm s}).
\end{equation}

The ratio $\eta_{{\rm s},ab}^{\alpha \beta}$ of signal
photon-number density $n_{{\rm s},ab}^{\alpha \beta,{\rm S}}$
emitted by surface SPDC and density $n_{{\rm s},ab}^{\alpha
\beta,{\rm V}}$ arising in volume SPDC,
\begin{equation}  
\eta_{{\rm s},ab}^{\alpha \beta}(\omega_{\rm s}) \equiv
\frac{n_{{\rm s},ab}^{\alpha \beta,S}(\omega_{\rm s})}{n_{{\rm
s},ab}^{\alpha \beta,{\rm V}}(\omega_{\rm s})}, \label{e72}
\end{equation}
provides insight into the nature of the whole SPDC process. For
the photon numbers, we define the ratio $R_{ab}^{\alpha \beta}$ of
number $N_{ab}^{\alpha \beta,{\rm S}}$ of photon pairs emitted at
the boundaries and number $N_{ab}^{\alpha \beta,{\rm V}}$ of
photon pairs created inside the layers:
\begin{equation}  
 R_{ab}^{\alpha \beta} \equiv  \frac{N_{ab}^{\alpha \beta,{\rm S}}}{N_{ab}^{\alpha \beta,{\rm V}}}.
\label{e71a}
\end{equation}

To reveal temporal characteristics of photon pairs, we define the
following spectral amplitude correlation function, that plays the
role of the usual spectral two-photon amplitude:
\begin{equation}  
 \phi_{ab}^{\alpha \beta}(\omega_{\rm s},\omega_{\rm i}) = \langle
 \mathrm{vac}\vert \hat{a}_{{\rm s}_a,\alpha}^{(\rm out)}(\omega_{\rm s})
 \hat{a}_{{\rm i}_b,\beta}^{(\rm out)}(\omega_{\rm i}) \vert \mathrm{vac}
 \rangle.
\end{equation}
Its Fourier transform provides us a temporal two-photon amplitude
$ \tilde{\phi}_{ab}^{\alpha \beta}(t_{\rm s},t_{\rm i}) $ that
gives the probability amplitude of detecting a signal photon
propagating in direction $a$ and polarized in direction $\alpha$
at time $t_{\rm s}$ together with its idler twin propagating in
direction $b$ with polarization $\beta$ at time $ t_{\rm i} $:
\begin{eqnarray} 
 \tilde{\phi}_{ab}^{\alpha \beta}(t_{\rm s},t_{\rm i}) &=& \int_{0}^{\infty}
  d\omega_{\rm s} \int_{0}^{\infty} d\omega_{\rm i}\, \phi_{ab}^{\alpha \beta}(\omega_{\rm s},\omega_{\rm i}) \nonumber \\
 & & \mbox{} \times \exp(-i\omega_{\rm s} t_{\rm s} -i\omega_{\rm i} t_{\rm i}).
 \label{e-two-photon-amp-time}
\end{eqnarray}
The corresponding normalized probability density $p_{ab}^{\alpha
\beta}$ is then obtained by the formula
\begin{equation} 
 p_{ab}^{\alpha \beta}(t_{\rm s},t_{\rm i}) = \frac{|\phi_{ab}^{\alpha \beta}(t_{\rm s},t_{\rm i})|^2}{\int_{-\infty}^{\infty}
  dt_{\rm s}' \int_{-\infty}^{\infty} dt_{\rm i}'
  |\phi_{ab}^{\alpha \beta}(t_{\rm s}',t_{\rm i}') |^2}.
 \label{e-joint-probability-detection-time}
\end{equation}
The temporal two-photon amplitude $ \tilde{\phi}_{ab}^{\alpha
\beta} $ also allows us to determine the normalized signal-field
photon flux $ p_{{\rm s},ab}^{\alpha \beta} $:
\begin{equation}   
 p_{{\rm s},ab}^{\alpha \beta}(t_{\rm s}) = \int_{-\infty}^{\infty} dt_{\rm i}\, p_{ab}^{\alpha \beta}(t_{\rm s},t_{\rm i}).
 \label{e-signal-probability-detection-time}
\end{equation}

\section{Properties of the emitted photon pairs}

In this section, we consider a typical nonlinear layered structure
made of alternating GaN/AlN layers under usual experimental
conditions. We assume a pump beam at central wavelength $
\lambda_{\rm p}^0 = 400 $~nm that impinges on the structure at
normal incidence from its left, is polarized along the ${\rm y}$
axis and has a Gaussian spectral profile:
\begin{equation}
 A_{{\rm p}_{\rm F},{\rm y}}^{(0)}(\omega_{\rm p}) = \sqrt{\sqrt{\frac{\mu_0}{\varepsilon_0 \pi}}
 \frac{E}{\pi \sigma_{\rm p}}} \exp\left[-\frac{(\omega_{\rm p} - \omega_{\rm p}^0)^2}{2\sigma_{\rm p}^2}\right].
\label{e74}
\end{equation}
The remaining input amplitudes $A_{{\rm p}_{\rm B},{\rm
y}}^{(N+1)}(\omega_{\rm p})$, $A_{{\rm p}_{\rm F},{\rm
x}}^{(0)}(\omega_{\rm p})$ and $A_{{\rm p}_{\rm B},{\rm
x}}^{(N+1)}(\omega_{\rm p})$ of the pump field are assumed to be
zero. In the analysis, the pump-beam energy $ E $ per unit area
equals $1 \times 10^{3}$~J/m$^2$ ($1$~mJ/mm$^2$) per one pulse.
The pump-beam spectral width $ \sigma_{\rm p} $ is set such that
the pump-beam intensity
spectral width (FWHM, full width in the
half of maximum) equals 7~nm. Assuming a transformed limited pump
pulse, its intensity temporal width equals 33~fs (FWHM). In the
analysis, we focus on the properties of photon pairs with both
photons propagating forward and the signal photon polarized along
the $ {\rm x} $-axis together with its idler twin polarized along
the $ {\rm y} $-axis. For simplicity, we omit both propagation and
polarization indices in the following discussion.

To obtain an efficient nonlinear layered structure, all three
interacting fields have to be sufficiently enhanced by
back-scattering inside the structure. This requires localization
of the fields into transmission peaks found near band gaps (for
details, \cite{PerinaJr2011}). This can be accomplished in two
steps. In the first step, layered structures with the pump beam
localized in a transmission peak near the band gap are identified.
Then, in the second step, the identified structures are analyzed
and those exhibiting the largest number of emitted photon pairs
are chosen.

\begin{figure}   
 \includegraphics[scale=1]{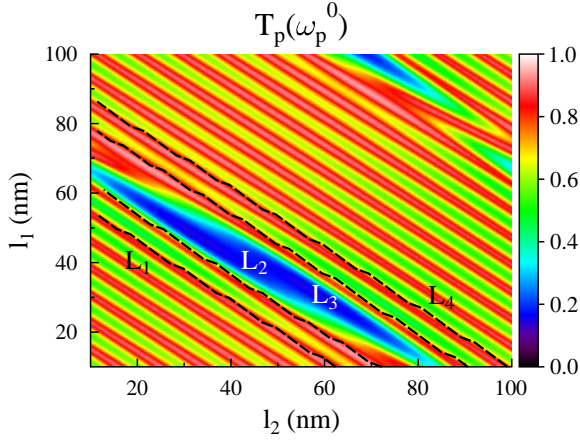}
 \caption{[Color online] Topo graph of linear intensity transmission coefficient
 $T_p$ as it depends on layers' lengths $l_1$ and $l_2$ for the pump
 beam at wavelength $\lambda_{\rm p}^0=400$~nm and structure having 10
 double layers GaN/AlN. Black dashed curves denoted as $L_1$,
 $L_2$, $L_3$, and $L_4$ indicate transmission peaks of the structures
 used in further analysis.}
\label{o-T_all}
\end{figure}

The considered structures were composed of 10 GaN and 10 AlN
mutually alternating layers. Their lengths $l_1$ (GaN) and $ l_2 $
(AlN) were assumed in interval $ (10\text{~nm},100\text{~nm})$, in
which the greatest enhancement of fields' amplitudes occurs. The
linear intensity transmission coefficient $ T_p $ for the pump
beam at central wavelength $\lambda_{\rm p}^0$ and for the
considered GaN/AlN structures is plotted in Fig.~\ref{o-T_all}. It
exhibits periodically alternating transmission and reflection
bands. In Fig.~\ref{o-T_all}, the band gaps are found in blue
regions whereas the transmission peaks are indicated by red
curves. Four transmission peaks indicated by black curves in
Fig.~\ref{o-T_all} are highlighted ($ L_j $, $ j=1,\ldots,4 $).
The peaks denoted as $L_2$ and $L_3$ occur next to a band gap and
so, according to the theory of band-gap structures, they provide
the greatest enhancement of pump-field amplitudes. For comparison,
we also analyze the structures lying in peaks $ L_1 $ and $ L_4 $.

\begin{figure}  
 \includegraphics[scale=1]{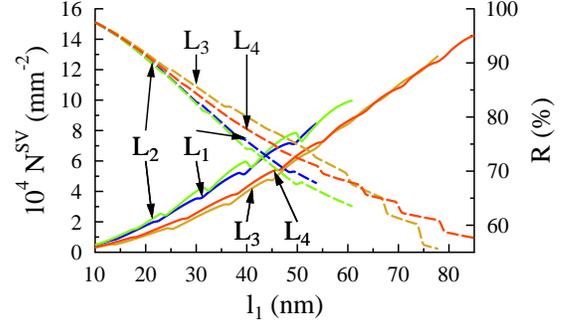}
 \caption{[Color online] Number $N^{\rm SV}$ of photon pairs emitted in both volume and surface SPDC (full curves)
 and ratio $R$ of photon-pair number arising in surface SPDC and that
 coming from volume SPDC (dashed curves) as they depend on GaN layers' length $
 l_1 $. The quantities are drawn for curves  $L_1 $ (blue curves), $ L_2 $ (green),
 $L_3$ (yellow) and $L_4$ (red) defined in
 Fig.~\ref{o-T_all}.}
\label{o-NsvR}
\end{figure}

The overall number $N^{\rm SV}$ of emitted photon pairs for the
structures lying on curves $L_1 $, $ L_2 $, $L_3$, and $L_4$
defined in the graph of Fig.~\ref{o-T_all} is determined in the
second step to find the most efficient structures. For all four
curves, the number $N^{\rm SV}$ of emitted photon pairs increases
with the increasing length $l_1$ of nonlinear GaN layers (see the
curves in Fig.~\ref{o-NsvR}). This increase originates in the
increasing amount of nonlinear GaN material inside the structure.
However, the observed dependence is nontrivial as interference of
the fields back-scattered inside the structure depends strongly on
the layers' lengths $ l_1 $ and $ l_2 $. The number $N^{\rm SV}$
of emitted photon pairs increases faster for the structures lying
on curves $L_1$ and $L_2$ situated below the band gap compared to
those found at curves $L_3$ and $L_4$ positioned above the band
gap. This is probably caused by the fact that the pump-field
amplitudes along the structure are localized preferably in the
nonlinear GaN layers for the peaks below the band gap, contrary to
the peaks above the band gap in which the pump-field amplitudes
are preferably localized in the linear AlN layers.

Contrary to the overall number $N^{\rm SV}$ of emitted photon
pairs, the ratio $R$ of photon-pair number $N^{\rm S}$ emitted at
the surfaces and number $N^{\rm V}$ of photon pairs created in the
volume decreases with the increasing length $ l_1 $ of GaN layers.
This is so as the number $N^{\rm S}$ of photon pairs emitted at
surfaces decreases with the increasing length $ l_1 $ and,
simultaneously, the number $N^{\rm V}$ of photon pairs generated
in the volume raises as the length $ l_1 $ increases. Decrease in
the ratio $R$ is faster for curves $L_1$ and $L_2$ as the number
$N^{\rm V}$ of photon pairs plotted as a function of length $ l_1
$ raises faster.

\begin{figure}  
 \includegraphics[scale=1]{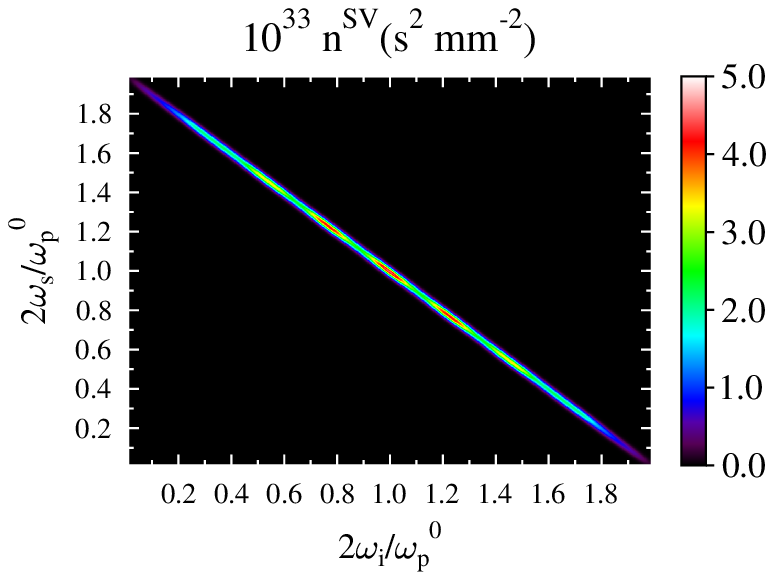}
 \centerline{(a)}
 \includegraphics[scale=1]{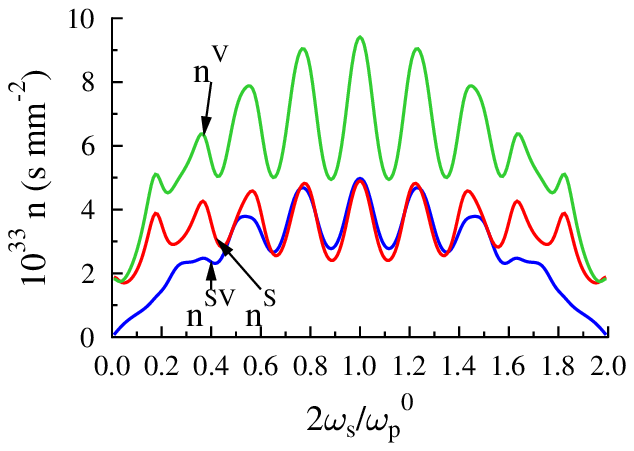}
 \centerline{(b)}
 \caption{[Color online] (a) Joint signal-idler spectral photon-number density $ n^{\rm SV}$
 of complete SPDC and (b) profiles of joint signal-idler spectral photon-number
 densities $n^{\rm S}$ (red curve), $n^{\rm V}$ (green) and $n^{\rm SV}$ (blue)
 arising in turn in surface, volume and
 complete SPDC along the line $ \omega_{\rm s} + \omega_{\rm i} = \omega_{\rm p}^0 $ as they depend
 on normalized signal frequency
 $\omega_{\rm s}/\omega_{\rm p}^0$; $ N = 20 $, $ l_1 = 60$~nm, $ l_2 =
 13$~nm, $ \lambda_{\rm p}^0 = 400$~nm.}
\label{fig5}
\end{figure}

For detailed analysis, we have chosen a structure composed of 10
GaN layers $l_1=60$~nm long and 10 AlN layers $l_2=13$~nm long.
Its joint signal-idler spectral photon-number density $n^{\rm SV}$
for the whole SPDC process is drawn in Fig.~\ref{fig5}(a).
Photon-pair emission occurs in the broad frequency range. Three
main peaks can be found in the spectral density $n^{\rm SV}$ of
both signal and idler fields shown in Fig.~\ref{fig5}(b).  The
main peaks are found at the central frequencies $2\omega_{\rm
s}/\omega_{\rm p}^0 = 2\omega_{\rm i}/\omega_{\rm p}^0=1 $ where $
n^{\rm SV} = 4.97 \times 10^{-33}$. On the other hand, the profile
of photon-number density $n^{\rm SV}$ plotted as a function of the
difference $ \omega_{\rm s} - \omega_{\rm i} $ of the signal and
idler frequencies is narrow as its spread is dominantly given by
the pump-field spectral width.

To get insight into the origin of photon pairs generated in SPDC,
we compare in parallel the contribution of volume SPDC and surface
SPCD to the complete SPDC process. These contributions are
compared in Fig.~\ref{fig5}(b) where the profiles of the
corresponding joint signal-idler spectral photon-number densities
$ n $ taken along the line $ \omega_{\rm s} + \omega_{\rm i} =
\omega_{\rm p}^0 $ are plotted. There occur nine resonant peaks in
the profiles $n^{\rm S}$ and $n^{\rm V}$ belonging to volume and
surface SPDC, respectively. Contrary to this, only five
well-recognized peaks are observed in the profile $n^{\rm SV}$
characterizing the complete SPDC process. This points out at
strong interference between the amplitudes describing volume and
surface SPDC processes. Indeed, this interference suppresses two
outermost peaks at both sides of the spectral profiles $n^{\rm S}$
and $n^{\rm V} $. The comparison of profiles in Fig.~\ref{fig5}(b)
for the densities $n^{\rm SV}$ and $n^{\rm V} $ identities volume
SPDC as being roughly twice intense compared to the complete SPDC
process. This means that surface SPDC has to be sufficiently
strong to cause the reduction of spectral photon-pair densities to
roughly one half via destructive interference. The profile of
density $n^{\rm S}$ created by surface SPDC and plotted in
Fig.~\ref{fig5}(b) confirms this reasoning. We note that the
profiles of all three densities $n^{\rm V}$, $n^{\rm S}$ and
$n^{\rm SV}$ cut along the line $ \omega_{\rm s} = \omega_{\rm i}
$ have comparable shapes resembling that of the pump-field
intensity spectrum.

\begin{figure} 
 \includegraphics[scale=1]{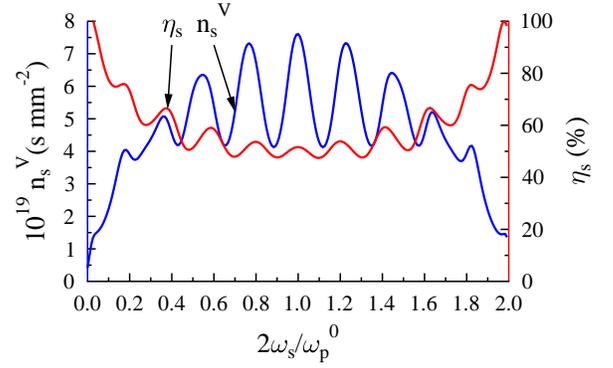}
 \caption{[Color online] Signal spectral photon-number density $n_{\rm s}^{\rm V}$ of volume SPDC (blue curve) and
  ratio $\eta_{\rm s}$ of the signal surface and volume photon-number densities (red curve) for the structure
  described in the caption to Fig.~5.}
\label{o-ns_rat}
\end{figure}

The relative contributions of surface and volume SPDC processes
are compared in Fig.~\ref{o-ns_rat} where the signal spectral
photon-number density  $n_{\rm s}^{\rm V}$ of volume SPDC and the
ratio $\eta_{\rm s}$ of the signal surface and volume spectral
photon-number densities are drawn. Volume SPDC is efficient in the
broad spectral range $ \omega_{\rm s} \in (0.1,0.9) \omega_{\rm
p}^0 $. The smallest values of ratio $\eta_s$ are reached in the
center of the emission interval ($ \eta_{\rm s} \approx 0.5 $)
where one surface photon pair is created together with about two
volume photon pairs. On the other hand, the values of ratio
$\eta_{\rm s}$ approach 1 at the edges of the spectral profile
$n_{\rm s}^{\rm V}$. This means that the volume and surface SPDC
processes are comparably strong in this region and the numbers of
emitted surface and volume photon pairs are comparable.

\begin{figure}  
 \includegraphics[scale=1]{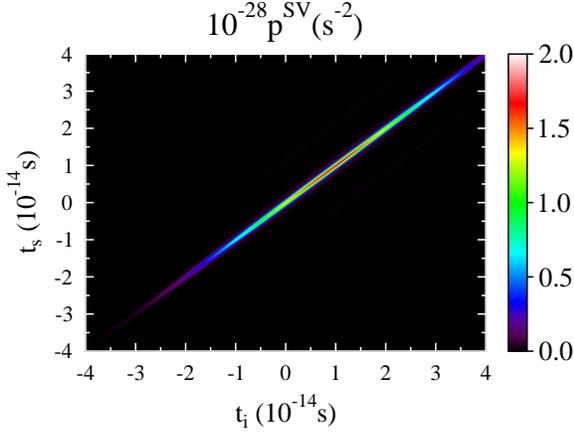}
 \caption{[Color online] Joint signal-idler probability density $p^{\rm SV}$ of
  the complete SPDC process as it depends on the signal- ($ t_{\rm s} $) and idler-photon
 ($ t_{\rm i} $) detection times.}
\label{o-phi_t}
\end{figure}

In time domain, the joint signal-idler probability densities
$p^{\rm V}$, $p^{\rm S}$ and $p^{\rm SV}$ of detecting a signal
photon at time $ t_{\rm s} $ and its idler twin at time $ t_{\rm
i} $ attain typical cigar shapes in their topo graphs in the $
t_{\rm s}-t_{\rm i} $ plane (for the probability density $p^{\rm
SV} $, see Fig.~\ref{o-phi_t}). The joint photon-number
probability densities $p^{\rm V}$ of volume SPDC and $p^{\rm S}$
of surface SPDC have similar profiles. Maximum of the probability
density $p^{\rm V}$ ($p^{\rm S}$) is reached at $ t\equiv t_{\rm
s}=t_{\rm i}=10.3$~fs ($ t=10.1$~fs). On the other hand, maximum
of the probability density $p^{\rm SV}$ is observed earlier, at $
t=9.5$~fs. This is the consequence of strong destructive
interference between the volume and surface contributions to SPDC
process. We note that this time gives relative average delay that
a signal (as well as an idler) photon needs to leave the structure
after being born 'inside' the propagating pump pulse.

\begin{figure}   
 \includegraphics[scale=1]{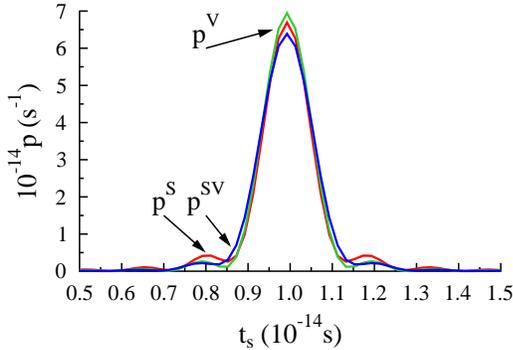}
 \caption{[Color online] Conditional probability densities $p^{\rm SV}$ (complete
 SPDC, blue curve), $p^{\rm V}$ (volume SPDC, green) and
 $p^{\rm S}$ (surface SPDC, red) of detecting a signal photon
 at time $ t_{\rm s} $ provided that its idler photon was detected at
 time $t_{\rm i}=10$~fs.}
\label{o-phi_t_cut}
\end{figure}

The profiles of conditional probabilities $p^{\rm S} $, $p^{\rm
V}$ and $p^{\rm SV}$ of detecting a signal photon at time $ t_{\rm
s} $ provided that its idler twin was detected at time $t_{\rm i}$
are close to each other. They are drawn for the analyzed structure
in Fig.~\ref{o-phi_t_cut} for $ t_{\rm i} = 10$~fs, where their
widths equal 1.4~fs (FWHM).

\begin{figure}  
 \includegraphics[scale=1]{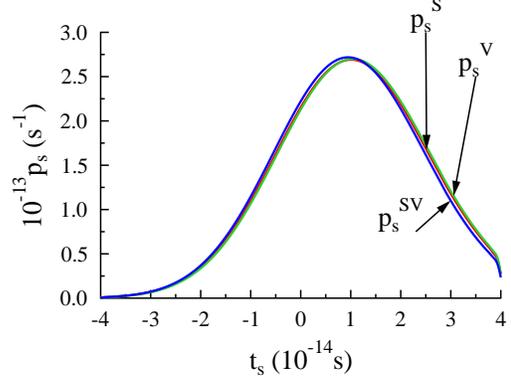}
 \caption{[Color online] Signal-field photon fluxes $p_{\rm s}^{\rm SV}$  (complete
 SPDC, blue curve), $p_{\rm s}^{\rm V}$ (volume SPDC, green) and $p_{\rm s}^{\rm S}$
 (surface SPDC, red).}
\label{o-phi_ts}
\end{figure}

Also the signal-field photon fluxes $p_s^{\rm S}$, $p_s^{\rm V}$
and $p_s^{\rm SV}$ are close to each other, as documented in
Fig.~\ref{o-phi_ts}. Their roughly Gaussian temporal profiles are
35.8~fs wide (FWHM), which is comparable to the pump-beam temporal
width.

In the analyzed structure, photons comprising the generated photon
pairs may leave the structure in both forward and backward
directions and also in different polarization combinations. The
total number $N^{\rm SV} $ of photon pairs leaving the structure
at both directions is $ 2.9 \times 10^{-3}$~mm$^{-2}$ per pulse.
Whereas the volume SPDC process would alone provide $N^{\rm V} =
6.9 \times 10^{-3}$~mm$^{-2}$ photon pairs per pulse, the surface
SPDC process alone would generate $N^{\rm S} = 4.1 \times
10^{-3}$~mm$^{-2}$ photon pairs per pulse. This means that the
efficiency of surface SPDC reaches around 60~\% of that of volume
SPDC. We note that the absolute photon-pair generation rates
reached in the analyzed structure are comparable in the magnitude
with those characterizing a perfectly phase-matched structure
containing the same amount of nonlinear GaN material as the
analyzed structure (for more details, see \cite{PerinaJr2011}).

\section{Conclusion}

The model of complete spontaneous parametric down-conversion
comprising both its volume and surface contributions has been
developed for 1D nonlinear layered structures considering
simultaneously the solution of Heisenberg equations in individual
layers and continuity requirements of the field's amplitudes at
the layers' boundaries. The analysis of fields' propagation around
the boundaries has allowed to clearly separate the volume and
surface contributions to the nonlinear process. Strong destructive
interference of the fields' amplitudes arising in the volume and
surface nonlinear processes has been observed. Owing to this
interference, the photon-pair generation rates equal around one
half of those that would be generated in the only volume nonlinear
process. The surface nonlinear process is in general weaker than
that in the volume, but both of them are comparably strong in the
spectral regions with lower efficiencies of photon-pair
generation.

\section{Acknowledgements}
The authors acknowledge project no.~15-08971S of GA\v{C}R and
project no. LO1305 of M\v{S}MT \v{C}R for support.


\begin{thebibliography}{43}%
\makeatletter
\providecommand \@ifxundefined [1]{%
 \@ifx{#1\undefined}
}%
\providecommand \@ifnum [1]{%
 \ifnum #1\expandafter \@firstoftwo
 \else \expandafter \@secondoftwo
 \fi
}%
\providecommand \@ifx [1]{%
 \ifx #1\expandafter \@firstoftwo
 \else \expandafter \@secondoftwo
 \fi
}%
\providecommand \natexlab [1]{#1}%
\providecommand \enquote  [1]{``#1''}%
\providecommand \bibnamefont  [1]{#1}%
\providecommand \bibfnamefont [1]{#1}%
\providecommand \citenamefont [1]{#1}%
\providecommand \href@noop [0]{\@secondoftwo}%
\providecommand \href [0]{\begingroup \@sanitize@url \@href}%
\providecommand \@href[1]{\@@startlink{#1}\@@href}%
\providecommand \@@href[1]{\endgroup#1\@@endlink}%
\providecommand \@sanitize@url [0]{\catcode `\\12\catcode
`\$12\catcode
  `\&12\catcode `\#12\catcode `\^12\catcode `\_12\catcode `\%12\relax}%
\providecommand \@@startlink[1]{}%
\providecommand \@@endlink[0]{}%
\providecommand \url  [0]{\begingroup\@sanitize@url \@url }%
\providecommand \@url [1]{\endgroup\@href {#1}{\urlprefix }}%
\providecommand \urlprefix  [0]{URL }%
\providecommand \Eprint [0]{\href }%
\providecommand \doibase [0]{http://dx.doi.org/}%
\providecommand \selectlanguage [0]{\@gobble}%
\providecommand \bibinfo  [0]{\@secondoftwo}%
\providecommand \bibfield  [0]{\@secondoftwo}%
\providecommand \translation [1]{[#1]}%
\providecommand \BibitemOpen [0]{}%
\providecommand \bibitemStop [0]{}%
\providecommand \bibitemNoStop [0]{.\EOS\space}%
\providecommand \EOS [0]{\spacefactor3000\relax}%
\providecommand \BibitemShut  [1]{\csname bibitem#1\endcsname}%
\let\auto@bib@innerbib\@empty
\bibitem [{\citenamefont {Louisell}\ \emph {et~al.}(1961)\citenamefont
  {Louisell}, \citenamefont {Yariv},\ and\ \citenamefont
  {Siegman}}]{Louisell1961}%
  \BibitemOpen
  \bibfield  {author} {\bibinfo {author} {\bibfnamefont {W.~H.}\ \bibnamefont
  {Louisell}}, \bibinfo {author} {\bibfnamefont {A.}~\bibnamefont {Yariv}}, \
  and\ \bibinfo {author} {\bibfnamefont {A.~E.}\ \bibnamefont {Siegman}},\
  }\bibfield  {title} {\enquote {\bibinfo {title} {Quantum fluctuations and
  noise in parametric processes. {I.}}}\ }\href@noop {} {\bibfield  {journal}
  {\bibinfo  {journal} {Phys. Rev.}\ }\textbf {\bibinfo {volume} {124}},\
  \bibinfo {pages} {1646--1654} (\bibinfo {year} {1961})}\BibitemShut {NoStop}%
\bibitem [{\citenamefont {Harris}\ \emph {et~al.}(1967)\citenamefont {Harris},
  \citenamefont {Oshman},\ and\ \citenamefont {Byer}}]{Harris1967}%
  \BibitemOpen
  \bibfield  {author} {\bibinfo {author} {\bibfnamefont {S.~E.}\ \bibnamefont
  {Harris}}, \bibinfo {author} {\bibfnamefont {M.~K.}\ \bibnamefont {Oshman}},
  \ and\ \bibinfo {author} {\bibfnamefont {R.~L.}\ \bibnamefont {Byer}},\
  }\bibfield  {title} {\enquote {\bibinfo {title} {Observation of tunable
  optical parametric fluorescence},}\ }\href@noop {} {\bibfield  {journal}
  {\bibinfo  {journal} {Phys. Rev. Lett.}\ }\textbf {\bibinfo {volume} {18}},\
  \bibinfo {pages} {732--734} (\bibinfo {year} {1967})}\BibitemShut {NoStop}%
\bibitem [{\citenamefont {Magde}\ and\ \citenamefont {Mahr}(1967)}]{Magde1967}%
  \BibitemOpen
  \bibfield  {author} {\bibinfo {author} {\bibfnamefont {D.}~\bibnamefont
  {Magde}}\ and\ \bibinfo {author} {\bibfnamefont {H.}~\bibnamefont {Mahr}},\
  }\bibfield  {title} {\enquote {\bibinfo {title} {Study in ammonium dihydrogen
  phosphate of spontaneous parametric interaction tunable from 4400 to {16
  000~\accent23 A}},}\ }\href@noop {} {\bibfield  {journal} {\bibinfo
  {journal} {Phys. Rev. Lett.}\ }\textbf {\bibinfo {volume} {18}},\ \bibinfo
  {pages} {905--907} (\bibinfo {year} {1967})}\BibitemShut {NoStop}%
\bibitem [{\citenamefont {Keller}\ and\ \citenamefont
  {Rubin}(1997)}]{Keller1997}%
  \BibitemOpen
  \bibfield  {author} {\bibinfo {author} {\bibfnamefont {T.~E.}\ \bibnamefont
  {Keller}}\ and\ \bibinfo {author} {\bibfnamefont {M.~H.}\ \bibnamefont
  {Rubin}},\ }\bibfield  {title} {\enquote {\bibinfo {title} {Theory of
  two-photon entanglement for spontaneous parametric down-conversion driven by
  a narrow pump pulse},}\ }\href@noop {} {\bibfield  {journal} {\bibinfo
  {journal} {Phys. Rev. A}\ }\textbf {\bibinfo {volume} {56}},\ \bibinfo
  {pages} {1534--1541} (\bibinfo {year} {1997})}\BibitemShut {NoStop}%
\bibitem [{\citenamefont {Svozil{\'\i}k}\ \emph {et~al.}(2012)\citenamefont
  {Svozil{\'\i}k}, \citenamefont {Pe{\v{r}}ina~Jr.},\ and\ \citenamefont
  {Torres}}]{Svozilik2012}%
  \BibitemOpen
  \bibfield  {author} {\bibinfo {author} {\bibfnamefont {J.}~\bibnamefont
  {Svozil{\'\i}k}}, \bibinfo {author} {\bibfnamefont {J.}~\bibnamefont
  {Pe{\v{r}}ina~Jr.}}, \ and\ \bibinfo {author} {\bibfnamefont {J.~P.}\
  \bibnamefont {Torres}},\ }\bibfield  {title} {\enquote {\bibinfo {title}
  {High spatial entanglement via chirped quasi-phase-matched optical parametric
  down-conversion},}\ }\href@noop {} {\bibfield  {journal} {\bibinfo  {journal}
  {Phys. Rev. A}\ }\textbf {\bibinfo {volume} {86}},\ \bibinfo {pages} {052318}
  (\bibinfo {year} {2012})}\BibitemShut {NoStop}%
\bibitem [{\citenamefont {{Grice}}\ \emph {et~al.}(2008)\citenamefont
  {{Grice}}, \citenamefont {{Bennink}}, \citenamefont {{Zhao}}, \citenamefont
  {{Meyer}}, \citenamefont {{Whitten}},\ and\ \citenamefont
  {{Shaw}}}]{Grice2008}%
  \BibitemOpen
  \bibfield  {author} {\bibinfo {author} {\bibfnamefont {W.~P.}\ \bibnamefont
  {{Grice}}}, \bibinfo {author} {\bibfnamefont {R.~S.}\ \bibnamefont
  {{Bennink}}}, \bibinfo {author} {\bibfnamefont {Z.}~\bibnamefont {{Zhao}}},
  \bibinfo {author} {\bibfnamefont {K.}~\bibnamefont {{Meyer}}}, \bibinfo
  {author} {\bibfnamefont {W.}~\bibnamefont {{Whitten}}}, \ and\ \bibinfo
  {author} {\bibfnamefont {R.}~\bibnamefont {{Shaw}}},\ }\bibfield  {title}
  {\enquote {\bibinfo {title} {Spectral and spatial effects in spontaneous
  parametric down-conversion with a focused pump},}\ }in\ \href@noop {} {\emph
  {\bibinfo {booktitle} {Quantum Communications and Quantum Imaging VI}}},\
  \bibinfo {series} {SPIE Conference Series}, Vol.\ \bibinfo {volume} {7092},\
  \bibinfo {editor} {edited by\ \bibinfo {editor} {\bibfnamefont {R.~E.}\
  \bibnamefont {Meyers}}, \bibinfo {editor} {\bibfnamefont {Y.}~\bibnamefont
  {Shih}}, \ and\ \bibinfo {editor} {\bibfnamefont {K.~S.}\ \bibnamefont
  {Deacon}}}\ (\bibinfo  {publisher} {SPIE, Bellingham},\ \bibinfo {year}
  {2008})\ p.\ \bibinfo {pages} {70920Q}\BibitemShut {NoStop}%
\bibitem [{\citenamefont {Jav\ifmmode~\mathring{u}\else \r{u}\fi{}rek}\ \emph
  {et~al.}(2014)\citenamefont {Jav\ifmmode~\mathring{u}\else \r{u}\fi{}rek},
  \citenamefont {Svozil\'{\i}k},\ and\ \citenamefont {Pe\ifmmode~\check{r}\else
  \v{r}\fi{}ina}}]{Javurek2014b}%
  \BibitemOpen
  \bibfield  {author} {\bibinfo {author} {\bibfnamefont {D.}~\bibnamefont
  {Jav\ifmmode~\mathring{u}\else \r{u}\fi{}rek}}, \bibinfo {author}
  {\bibfnamefont {J.}~\bibnamefont {Svozil\'{\i}k}}, \ and\ \bibinfo {author}
  {\bibfnamefont {J.}~\bibnamefont {Pe\ifmmode~\check{r}\else \v{r}\fi{}ina}},\
  }\bibfield  {title} {\enquote {\bibinfo {title} {Emission of
  orbital-angular-momentum-entangled photon pairs in a nonlinear ring fiber
  utilizing spontaneous parametric down-conversion},}\ }\href {\doibase
  10.1103/PhysRevA.90.043844} {\bibfield  {journal} {\bibinfo  {journal} {Phys.
  Rev. A}\ }\textbf {\bibinfo {volume} {90}},\ \bibinfo {pages} {043844}
  (\bibinfo {year} {2014})}\BibitemShut {NoStop}%
\bibitem [{\citenamefont {{Pe\v{r}ina~Jr.}}\ \emph
  {et~al.}(2009{\natexlab{a}})\citenamefont {{Pe\v{r}ina~Jr.}}, \citenamefont
  {Luk\v{s}}, \citenamefont {Haderka},\ and\ \citenamefont
  {Scalora}}]{PerinaJr2009}%
  \BibitemOpen
  \bibfield  {author} {\bibinfo {author} {\bibfnamefont {J.}~\bibnamefont
  {{Pe\v{r}ina~Jr.}}}, \bibinfo {author} {\bibfnamefont {A.}~\bibnamefont
  {Luk\v{s}}}, \bibinfo {author} {\bibfnamefont {O.}~\bibnamefont {Haderka}}, \
  and\ \bibinfo {author} {\bibfnamefont {M.}~\bibnamefont {Scalora}},\
  }\bibfield  {title} {\enquote {\bibinfo {title} {Surface spontaneous
  parametric down-conversion},}\ }\href@noop {} {\bibfield  {journal} {\bibinfo
   {journal} {Phys. Rev. Lett.}\ }\textbf {\bibinfo {volume} {103}},\ \bibinfo
  {pages} {063902} (\bibinfo {year} {2009}{\natexlab{a}})}\BibitemShut
  {NoStop}%
\bibitem [{\citenamefont {{Pe\v{r}ina~Jr.}}(2014)}]{PerinaJr2014}%
  \BibitemOpen
  \bibfield  {author} {\bibinfo {author} {\bibfnamefont {J.}~\bibnamefont
  {{Pe\v{r}ina~Jr.}}},\ }\bibfield  {title} {\enquote {\bibinfo {title}
  {Spontaneous parametric down-conversion in nonlinear layered media},}\ }in\
  \href@noop {} {\emph {\bibinfo {booktitle} {Progress in Optics}}},\
  Vol.~\bibinfo {volume} {59},\ \bibinfo {editor} {edited by\ \bibinfo {editor}
  {\bibfnamefont {E.}~\bibnamefont {Wolf}}}\ (\bibinfo  {publisher} {Elsevier,
  Amsterdam},\ \bibinfo {year} {2014})\ pp.\ \bibinfo {pages}
  {89--158}\BibitemShut {NoStop}%
\bibitem [{\citenamefont {Boyd}(2003)}]{Boyd2003}%
  \BibitemOpen
  \bibfield  {author} {\bibinfo {author} {\bibfnamefont {R.~W.}\ \bibnamefont
  {Boyd}},\ }\href@noop {} {\emph {\bibinfo {title} {Nonlinear Optics, 2nd
  edition}}}\ (\bibinfo  {publisher} {Academic Press, New York},\ \bibinfo
  {year} {2003})\BibitemShut {NoStop}%
\bibitem [{\citenamefont {Dmitriev}\ \emph {et~al.}(1999)\citenamefont
  {Dmitriev}, \citenamefont {Gurzadyan},\ and\ \citenamefont
  {Nikogosyan}}]{Dmitriev1999}%
  \BibitemOpen
  \bibfield  {author} {\bibinfo {author} {\bibfnamefont {V.~G.}\ \bibnamefont
  {Dmitriev}}, \bibinfo {author} {\bibfnamefont {G.~G.}\ \bibnamefont
  {Gurzadyan}}, \ and\ \bibinfo {author} {\bibfnamefont {D.~N.}\ \bibnamefont
  {Nikogosyan}},\ }\href@noop {} {\emph {\bibinfo {title} {Handbook of
  Nonlinear Optical Crystals}}}\ (\bibinfo  {publisher} {Springer-Verlag,
  Berlin Heidelberg},\ \bibinfo {year} {1999})\BibitemShut {NoStop}%
\bibitem [{\citenamefont {{Pe\v{r}ina~Jr.}}\ \emph {et~al.}(2006)\citenamefont
  {{Pe\v{r}ina~Jr.}}, \citenamefont {Centini}, \citenamefont {Sibilia},
  \citenamefont {Bertolotti},\ and\ \citenamefont {Scalora}}]{PerinaJr2006}%
  \BibitemOpen
  \bibfield  {author} {\bibinfo {author} {\bibfnamefont {J.}~\bibnamefont
  {{Pe\v{r}ina~Jr.}}}, \bibinfo {author} {\bibfnamefont {M.}~\bibnamefont
  {Centini}}, \bibinfo {author} {\bibfnamefont {C.}~\bibnamefont {Sibilia}},
  \bibinfo {author} {\bibfnamefont {M.}~\bibnamefont {Bertolotti}}, \ and\
  \bibinfo {author} {\bibfnamefont {M.}~\bibnamefont {Scalora}},\ }\bibfield
  {title} {\enquote {\bibinfo {title} {Properties of entangled photon pairs
  generated in one-dimensional nonlinear photonic-band-gap structures},}\
  }\href@noop {} {\bibfield  {journal} {\bibinfo  {journal} {Phys. Rev. A}\
  }\textbf {\bibinfo {volume} {73}},\ \bibinfo {eid} {033823} (\bibinfo {year}
  {2006})}\BibitemShut {NoStop}%
\bibitem [{\citenamefont {{Pe\v{r}ina~Jr.}}\ \emph
  {et~al.}(2009{\natexlab{b}})\citenamefont {{Pe\v{r}ina~Jr.}}, \citenamefont
  {Centini}, \citenamefont {Sibilia},\ and\ \citenamefont
  {Bertolotti}}]{PerinaJr2009c}%
  \BibitemOpen
  \bibfield  {author} {\bibinfo {author} {\bibfnamefont {J.}~\bibnamefont
  {{Pe\v{r}ina~Jr.}}}, \bibinfo {author} {\bibfnamefont {M.}~\bibnamefont
  {Centini}}, \bibinfo {author} {\bibfnamefont {C.}~\bibnamefont {Sibilia}}, \
  and\ \bibinfo {author} {\bibfnamefont {M.}~\bibnamefont {Bertolotti}},\
  }\bibfield  {title} {\enquote {\bibinfo {title} {Photon-pair generation in
  random nonlinear layered structures},}\ }\href@noop {} {\bibfield  {journal}
  {\bibinfo  {journal} {Phys. Rev. A}\ }\textbf {\bibinfo {volume} {80}},\
  \bibinfo {pages} {033844} (\bibinfo {year} {2009}{\natexlab{b}})}\BibitemShut
  {NoStop}%
\bibitem [{\citenamefont {{Pe\v{r}ina~Jr.}}(2011)}]{PerinaJr2011}%
  \BibitemOpen
  \bibfield  {author} {\bibinfo {author} {\bibfnamefont {J.}~\bibnamefont
  {{Pe\v{r}ina~Jr.}}},\ }\bibfield  {title} {\enquote {\bibinfo {title}
  {Spatial properties of entangled photon pairs generated in nonlinear layered
  structures},}\ }\href@noop {} {\bibfield  {journal} {\bibinfo  {journal}
  {Phys. Rev. A}\ }\textbf {\bibinfo {volume} {84}},\ \bibinfo {pages} {053840}
  (\bibinfo {year} {2011})}\BibitemShut {NoStop}%
\bibitem [{\citenamefont {{Jav\accent23 urek}}\ \emph
  {et~al.}(2012)\citenamefont {{Jav\accent23 urek}}, \citenamefont
  {Svozil\'{\i}k},\ and\ \citenamefont {{Pe\v{r}ina~Jr.}}}]{Javurek2012}%
  \BibitemOpen
  \bibfield  {author} {\bibinfo {author} {\bibfnamefont {D.}~\bibnamefont
  {{Jav\accent23 urek}}}, \bibinfo {author} {\bibfnamefont {J.}~\bibnamefont
  {Svozil\'{\i}k}}, \ and\ \bibinfo {author} {\bibfnamefont {J.}~\bibnamefont
  {{Pe\v{r}ina~Jr.}}},\ }\bibfield  {title} {\enquote {\bibinfo {title}
  {{Entangled photon-pair generation in metallo-dielectric photonic bandgap
  structures}},}\ }in\ \href@noop {} {\emph {\bibinfo {booktitle} {Wave and
  Quantum Aspects of Contemporary Optics}}},\ \bibinfo {series} {SPIE
  Conference proceedings}, Vol.\ \bibinfo {volume} {8697},\ \bibinfo {editor}
  {edited by\ \bibinfo {editor} {\bibfnamefont {J.}~\bibnamefont
  {{Pe\v{r}ina~Jr.}}}, \bibinfo {editor} {\bibfnamefont {L.}~\bibnamefont
  {No\v{z}ka}}, \bibinfo {editor} {\bibfnamefont {M.}~\bibnamefont
  {Hrabovsk\'{y}}}, \bibinfo {editor} {\bibfnamefont {D.}~\bibnamefont
  {Sender\'{a}kov\'{a}}}, \bibinfo {editor} {\bibfnamefont {W.}~\bibnamefont
  {Urbanczyk}}, \ and\ \bibinfo {editor} {\bibfnamefont {O.}~\bibnamefont
  {Haderka}}}\ (\bibinfo  {publisher} {SPIE, Bellingham},\ \bibinfo {year}
  {2012})\BibitemShut {NoStop}%
\bibitem [{\citenamefont {{Jav\accent23 urek}}\ \emph
  {et~al.}(2014)\citenamefont {{Jav\accent23 urek}}, \citenamefont
  {Svozil\'{\i}k},\ and\ \citenamefont {{Pe\v{r}ina~Jr.}}}]{Javurek2014x}%
  \BibitemOpen
  \bibfield  {author} {\bibinfo {author} {\bibfnamefont {D.}~\bibnamefont
  {{Jav\accent23 urek}}}, \bibinfo {author} {\bibfnamefont {J.}~\bibnamefont
  {Svozil\'{\i}k}}, \ and\ \bibinfo {author} {\bibfnamefont {J.}~\bibnamefont
  {{Pe\v{r}ina~Jr.}}},\ }\bibfield  {title} {\enquote {\bibinfo {title}
  {{Spontaneous parametric down conversion in nonlinear metal-dielectric
  layered media}},}\ }in\ \href@noop {} {\emph {\bibinfo {booktitle} {Wave and
  Quantum Aspects of Contemporary Optics}}},\ \bibinfo {series} {SPIE
  Conference proceedings}, Vol.\ \bibinfo {volume} {9441},\ \bibinfo {editor}
  {edited by\ \bibinfo {editor} {\bibfnamefont {A.}~\bibnamefont
  {Popio³ek-Masajada}}\ and\ \bibinfo {editor} {\bibfnamefont
  {W.}~\bibnamefont {Urba\'{n}czyk}}}\ (\bibinfo  {publisher} {SPIE,
  Bellingham},\ \bibinfo {year} {2014})\ p.\ \bibinfo {pages}
  {94410V}\BibitemShut {NoStop}%
\bibitem [{\citenamefont {Zhu}\ \emph {et~al.}(2012)\citenamefont {Zhu},
  \citenamefont {Tang}, \citenamefont {Qian}, \citenamefont {Helt},
  \citenamefont {Liscidini}, \citenamefont {Sipe}, \citenamefont {Corbari},
  \citenamefont {Canagasabey}, \citenamefont {Ibsen},\ and\ \citenamefont
  {Kazansky}}]{Zhu2012}%
  \BibitemOpen
  \bibfield  {author} {\bibinfo {author} {\bibfnamefont {E.~Y.}\ \bibnamefont
  {Zhu}}, \bibinfo {author} {\bibfnamefont {Z.}~\bibnamefont {Tang}}, \bibinfo
  {author} {\bibfnamefont {L.}~\bibnamefont {Qian}}, \bibinfo {author}
  {\bibfnamefont {L.~G.}\ \bibnamefont {Helt}}, \bibinfo {author}
  {\bibfnamefont {M.}~\bibnamefont {Liscidini}}, \bibinfo {author}
  {\bibfnamefont {J.~E.}\ \bibnamefont {Sipe}}, \bibinfo {author}
  {\bibfnamefont {C.}~\bibnamefont {Corbari}}, \bibinfo {author} {\bibfnamefont
  {A.}~\bibnamefont {Canagasabey}}, \bibinfo {author} {\bibfnamefont
  {M.}~\bibnamefont {Ibsen}}, \ and\ \bibinfo {author} {\bibfnamefont {P.~G.}\
  \bibnamefont {Kazansky}},\ }\bibfield  {title} {\enquote {\bibinfo {title}
  {Direct generation of polarization-entangled photon pairs in a poled
  fiber},}\ }\href@noop {} {\bibfield  {journal} {\bibinfo  {journal} {Phys.
  Rev. Lett.}\ }\textbf {\bibinfo {volume} {108}},\ \bibinfo {pages} {213902}
  (\bibinfo {year} {2012})}\BibitemShut {NoStop}%
\bibitem [{\citenamefont {Jav\r{u}rek}\ \emph
  {et~al.}(2014{\natexlab{a}})\citenamefont {Jav\r{u}rek}, \citenamefont
  {Svozil\'{i}k},\ and\ \citenamefont {Pe\v{r}ina~Jr.}}]{Javurek2014a}%
  \BibitemOpen
  \bibfield  {author} {\bibinfo {author} {\bibfnamefont {D.}~\bibnamefont
  {Jav\r{u}rek}}, \bibinfo {author} {\bibfnamefont {J.}~\bibnamefont
  {Svozil\'{i}k}}, \ and\ \bibinfo {author} {\bibfnamefont {J.}~\bibnamefont
  {Pe\v{r}ina~Jr.}},\ }\bibfield  {title} {\enquote {\bibinfo {title} {Proposal
  for the generation of photon pairs with nonzero orbital angular momentum in a
  ring fiber},}\ }\href {\doibase 10.1364/OE.22.023743} {\bibfield  {journal}
  {\bibinfo  {journal} {Opt. Express}\ }\textbf {\bibinfo {volume} {22}},\
  \bibinfo {pages} {23743--23748} (\bibinfo {year}
  {2014}{\natexlab{a}})}\BibitemShut {NoStop}%
\bibitem [{\citenamefont {Eckstein}\ \emph {et~al.}(2011)\citenamefont
  {Eckstein}, \citenamefont {Christ}, \citenamefont {Mosley},\ and\
  \citenamefont {Silberhorn}}]{Eckstein2011}%
  \BibitemOpen
  \bibfield  {author} {\bibinfo {author} {\bibfnamefont {A.}~\bibnamefont
  {Eckstein}}, \bibinfo {author} {\bibfnamefont {A.}~\bibnamefont {Christ}},
  \bibinfo {author} {\bibfnamefont {P.~J.}\ \bibnamefont {Mosley}}, \ and\
  \bibinfo {author} {\bibfnamefont {C.}~\bibnamefont {Silberhorn}},\ }\bibfield
   {title} {\enquote {\bibinfo {title} {Highly efficient single-pass source of
  pulsed single-mode twin beams of light},}\ }\href@noop {} {\bibfield
  {journal} {\bibinfo  {journal} {Phys. Rev. Lett.}\ }\textbf {\bibinfo
  {volume} {106}},\ \bibinfo {pages} {013603} (\bibinfo {year}
  {2011})}\BibitemShut {NoStop}%
\bibitem [{\citenamefont {Jachura}\ \emph {et~al.}(2014)\citenamefont
  {Jachura}, \citenamefont {Karpinski}, \citenamefont {Radzewicz},\ and\
  \citenamefont {Banaszek}}]{Jachura2014}%
  \BibitemOpen
  \bibfield  {author} {\bibinfo {author} {\bibfnamefont {M.}~\bibnamefont
  {Jachura}}, \bibinfo {author} {\bibfnamefont {M.}~\bibnamefont {Karpinski}},
  \bibinfo {author} {\bibfnamefont {C.}~\bibnamefont {Radzewicz}}, \ and\
  \bibinfo {author} {\bibfnamefont {K.}~\bibnamefont {Banaszek}},\ }\bibfield
  {title} {\enquote {\bibinfo {title} {High-visibility nonclassical
  interference of photon pairs generated in a multimode nonlinear waveguide},}\
  }\href@noop {} {\bibfield  {journal} {\bibinfo  {journal} {Opt. Express}\
  }\textbf {\bibinfo {volume} {22}},\ \bibinfo {pages} {8624---8632} (\bibinfo
  {year} {2014})}\BibitemShut {NoStop}%
\bibitem [{\citenamefont {Machulka}\ \emph {et~al.}(2013)\citenamefont
  {Machulka}, \citenamefont {Svozil\'{\i}k}, \citenamefont {Soubusta},
  \citenamefont {{Pe\v{r}ina Jr}.},\ and\ \citenamefont
  {Haderka}}]{Machulka2013}%
  \BibitemOpen
  \bibfield  {author} {\bibinfo {author} {\bibfnamefont {R.}~\bibnamefont
  {Machulka}}, \bibinfo {author} {\bibfnamefont {J.}~\bibnamefont
  {Svozil\'{\i}k}}, \bibinfo {author} {\bibfnamefont {J.}~\bibnamefont
  {Soubusta}}, \bibinfo {author} {\bibfnamefont {J.}~\bibnamefont {{Pe\v{r}ina
  Jr}.}}, \ and\ \bibinfo {author} {\bibfnamefont {O.}~\bibnamefont
  {Haderka}},\ }\bibfield  {title} {\enquote {\bibinfo {title} {Spatial and
  spectral properties of the pulsed second-harmonic generation in a {PP-KTP}
  waveguide},}\ }\href@noop {} {\bibfield  {journal} {\bibinfo  {journal}
  {Phys. Rev. A}\ }\textbf {\bibinfo {volume} {87}},\ \bibinfo {eid} {013836}
  (\bibinfo {year} {2013})}\BibitemShut {NoStop}%
\bibitem [{\citenamefont {Clausen}\ \emph {et~al.}(2014)\citenamefont
  {Clausen}, \citenamefont {Bussieres}, \citenamefont {Tiranov}, \citenamefont
  {Herrmann}, \citenamefont {Silberhorn}, \citenamefont {Sohler}, \citenamefont
  {Afzelius},\ and\ \citenamefont {Gisin}}]{Clausen2014}%
  \BibitemOpen
  \bibfield  {author} {\bibinfo {author} {\bibfnamefont {C.}~\bibnamefont
  {Clausen}}, \bibinfo {author} {\bibfnamefont {F.}~\bibnamefont {Bussieres}},
  \bibinfo {author} {\bibfnamefont {A.}~\bibnamefont {Tiranov}}, \bibinfo
  {author} {\bibfnamefont {H.}~\bibnamefont {Herrmann}}, \bibinfo {author}
  {\bibfnamefont {C.}~\bibnamefont {Silberhorn}}, \bibinfo {author}
  {\bibfnamefont {W.}~\bibnamefont {Sohler}}, \bibinfo {author} {\bibfnamefont
  {M.}~\bibnamefont {Afzelius}}, \ and\ \bibinfo {author} {\bibfnamefont
  {N.}~\bibnamefont {Gisin}},\ }\bibfield  {title} {\enquote {\bibinfo {title}
  {A source of polarization-entangled photon pairs interfacing quantum memories
  with telecom photons},}\ }\href@noop {} {\bibfield  {journal} {\bibinfo
  {journal} {N. J. Phys.}\ }\textbf {\bibinfo {volume} {16}} (\bibinfo {year}
  {2014})}\BibitemShut {NoStop}%
\bibitem [{\citenamefont {Chen}\ \emph {et~al.}(2014)\citenamefont {Chen},
  \citenamefont {Xu}, \citenamefont {Bai}, \citenamefont {Luo}, \citenamefont
  {Zhong}, \citenamefont {Dai}, \citenamefont {Lu},\ and\ \citenamefont
  {Zhu}}]{Chen2014}%
  \BibitemOpen
  \bibfield  {author} {\bibinfo {author} {\bibfnamefont {L.}~\bibnamefont
  {Chen}}, \bibinfo {author} {\bibfnamefont {P.}~\bibnamefont {Xu}}, \bibinfo
  {author} {\bibfnamefont {Y.~F.}\ \bibnamefont {Bai}}, \bibinfo {author}
  {\bibfnamefont {X.~W.}\ \bibnamefont {Luo}}, \bibinfo {author} {\bibfnamefont
  {M.~L.}\ \bibnamefont {Zhong}}, \bibinfo {author} {\bibfnamefont
  {M.}~\bibnamefont {Dai}}, \bibinfo {author} {\bibfnamefont {M.~H.}\
  \bibnamefont {Lu}}, \ and\ \bibinfo {author} {\bibfnamefont {S.~N.}\
  \bibnamefont {Zhu}},\ }\bibfield  {title} {\enquote {\bibinfo {title}
  {Concurrent optical parametric down-conversion in $\chi^{(2)}$ nonlinear
  photonic crystals},}\ }\href {\doibase 10.1364/OE.22.013164} {\bibfield
  {journal} {\bibinfo  {journal} {Opt. Express}\ }\textbf {\bibinfo {volume}
  {22}},\ \bibinfo {pages} {13164--13169} (\bibinfo {year} {2014})}\BibitemShut
  {NoStop}%
\bibitem [{\citenamefont {Hayata}\ and\ \citenamefont
  {Koshiba}(1991)}]{Hayata1991}%
  \BibitemOpen
  \bibfield  {author} {\bibinfo {author} {\bibfnamefont {K}~\bibnamefont
  {Hayata}}\ and\ \bibinfo {author} {\bibfnamefont {M}~\bibnamefont
  {Koshiba}},\ }\bibfield  {title} {\enquote {\bibinfo {title}
  {Quasi-phase-matched multiwave mixing in a periodically poled ferroelectric
  crystal},}\ }\href {\doibase 10.1364/OL.16.000560} {\bibfield  {journal}
  {\bibinfo  {journal} {Opt. Lett.}\ }\textbf {\bibinfo {volume} {16}},\
  \bibinfo {pages} {560--562} (\bibinfo {year} {1991})}\BibitemShut {NoStop}%
\bibitem [{\citenamefont {Lim}\ \emph {et~al.}(1989)\citenamefont {Lim},
  \citenamefont {Fejer},\ and\ \citenamefont {Byer}}]{Lim1989}%
  \BibitemOpen
  \bibfield  {author} {\bibinfo {author} {\bibfnamefont {E.~J.}\ \bibnamefont
  {Lim}}, \bibinfo {author} {\bibfnamefont {M.~M.}\ \bibnamefont {Fejer}}, \
  and\ \bibinfo {author} {\bibfnamefont {R.~L.}\ \bibnamefont {Byer}},\
  }\bibfield  {title} {\enquote {\bibinfo {title} {2nd-harmonic generation of
  green light in periodically poled planar lithium-niobate wave-guide},}\
  }\href {\doibase 10.1049/el:19890127} {\bibfield  {journal} {\bibinfo
  {journal} {Electron. Lett.}\ }\textbf {\bibinfo {volume} {25}},\ \bibinfo
  {pages} {174--175} (\bibinfo {year} {1989})}\BibitemShut {NoStop}%
\bibitem [{\citenamefont {Shinozaki}\ \emph {et~al.}(1992)\citenamefont
  {Shinozaki}, \citenamefont {Fukunaga}, \citenamefont {Watanabe},\ and\
  \citenamefont {Kamijoh}}]{Shinozaki1992}%
  \BibitemOpen
  \bibfield  {author} {\bibinfo {author} {\bibfnamefont {K.}~\bibnamefont
  {Shinozaki}}, \bibinfo {author} {\bibfnamefont {T.}~\bibnamefont {Fukunaga}},
  \bibinfo {author} {\bibfnamefont {K.}~\bibnamefont {Watanabe}}, \ and\
  \bibinfo {author} {\bibfnamefont {T.}~\bibnamefont {Kamijoh}},\ }\bibfield
  {title} {\enquote {\bibinfo {title} {Automatic quasiphase matching for
  2nd-harmonic generation in a periodically poled {LiNbO${}_3$} wave-guide},}\
  }\href {\doibase {10.1063/1.350748}} {\bibfield  {journal} {\bibinfo
  {journal} {J. Appl. Phys.}\ }\textbf {\bibinfo {volume} {71}},\ \bibinfo
  {pages} {22--27} (\bibinfo {year} {1992})}\BibitemShut {NoStop}%
\bibitem [{\citenamefont {Kashyap}(1991)}]{Kashyap1991}%
  \BibitemOpen
  \bibfield  {author} {\bibinfo {author} {\bibfnamefont {R.}~\bibnamefont
  {Kashyap}},\ }\bibfield  {title} {\enquote {\bibinfo {title} {Phase-matched
  2nd-harmonic generation in periodically poled optical fibers},}\ }\href
  {\doibase 10.1063/1.104372} {\bibfield  {journal} {\bibinfo  {journal} {Appl.
  Phys. Lett.}\ }\textbf {\bibinfo {volume} {58}},\ \bibinfo {pages}
  {1233--1235} (\bibinfo {year} {1991})}\BibitemShut {NoStop}%
\bibitem [{\citenamefont {Chmela}(1991)}]{Chmela1991}%
  \BibitemOpen
  \bibfield  {author} {\bibinfo {author} {\bibfnamefont {P.}~\bibnamefont
  {Chmela}},\ }\bibfield  {title} {\enquote {\bibinfo {title} {Preparation of
  optical fibers for effective 2nd-harmonic generation by the poling
  technique},}\ }\href {\doibase {10.1364/OL.16.000443}} {\bibfield  {journal}
  {\bibinfo  {journal} {Opt. Lett.}\ }\textbf {\bibinfo {volume} {16}},\
  \bibinfo {pages} {443--445} (\bibinfo {year} {1991})}\BibitemShut {NoStop}%
\bibitem [{\citenamefont {Harris}(2007)}]{Harris2007}%
  \BibitemOpen
  \bibfield  {author} {\bibinfo {author} {\bibfnamefont {S.~E.}\ \bibnamefont
  {Harris}},\ }\bibfield  {title} {\enquote {\bibinfo {title} {Chirp and
  compress: Toward single-cycle biphotons},}\ }\href@noop {} {\bibfield
  {journal} {\bibinfo  {journal} {Phys. Rev. Lett.}\ }\textbf {\bibinfo
  {volume} {98}},\ \bibinfo {pages} {063602} (\bibinfo {year}
  {2007})}\BibitemShut {NoStop}%
\bibitem [{\citenamefont {Brida}\ \emph {et~al.}(2009)\citenamefont {Brida},
  \citenamefont {Chekhova}, \citenamefont {Degiovanni}, \citenamefont
  {Genovese}, \citenamefont {Kitaeva}, \citenamefont {Meda},\ and\
  \citenamefont {Shumilkina}}]{Brida2009}%
  \BibitemOpen
  \bibfield  {author} {\bibinfo {author} {\bibfnamefont {G.}~\bibnamefont
  {Brida}}, \bibinfo {author} {\bibfnamefont {M.~V.}\ \bibnamefont {Chekhova}},
  \bibinfo {author} {\bibfnamefont {I.~P.}\ \bibnamefont {Degiovanni}},
  \bibinfo {author} {\bibfnamefont {M.}~\bibnamefont {Genovese}}, \bibinfo
  {author} {\bibfnamefont {G.~Kh.}\ \bibnamefont {Kitaeva}}, \bibinfo {author}
  {\bibfnamefont {A.}~\bibnamefont {Meda}}, \ and\ \bibinfo {author}
  {\bibfnamefont {O.~A.}\ \bibnamefont {Shumilkina}},\ }\bibfield  {title}
  {\enquote {\bibinfo {title} {Chirped biphotons and their compression in
  optical fibers},}\ }\href@noop {} {\bibfield  {journal} {\bibinfo  {journal}
  {Phys. Rev. Lett.}\ }\textbf {\bibinfo {volume} {103}},\ \bibinfo {pages}
  {193602} (\bibinfo {year} {2009})}\BibitemShut {NoStop}%
\bibitem [{\citenamefont {Svozil\'{\i}k}\ and\ \citenamefont
  {{Pe\v{r}ina~Jr.}}(2009)}]{Svozilik2009}%
  \BibitemOpen
  \bibfield  {author} {\bibinfo {author} {\bibfnamefont {J.}~\bibnamefont
  {Svozil\'{\i}k}}\ and\ \bibinfo {author} {\bibfnamefont {J.}~\bibnamefont
  {{Pe\v{r}ina~Jr.}}},\ }\bibfield  {title} {\enquote {\bibinfo {title}
  {Properties of entangled photon pairs generated in periodically poled
  nonlinear crystals},}\ }\href@noop {} {\bibfield  {journal} {\bibinfo
  {journal} {Phys. Rev. A}\ }\textbf {\bibinfo {volume} {80}},\ \bibinfo
  {pages} {023819} (\bibinfo {year} {2009})}\BibitemShut {NoStop}%
\bibitem [{\citenamefont {Svozil\'ik}\ and\ \citenamefont
  {{Pe\v{r}ina~Jr.}}(2011)}]{Svozilik2011a}%
  \BibitemOpen
  \bibfield  {author} {\bibinfo {author} {\bibfnamefont {J.}~\bibnamefont
  {Svozil\'ik}}\ and\ \bibinfo {author} {\bibfnamefont {J.}~\bibnamefont
  {{Pe\v{r}ina~Jr.}}},\ }\bibfield  {title} {\enquote {\bibinfo {title}
  {{Intense ultra-broadband down-conversion from randomly poled nonlinear
  crystals}},}\ }in\ \href@noop {} {\emph {\bibinfo {booktitle} {Nonlinear
  Optics and Applications V}}},\ \bibinfo {series} {SPIE Conference
  proceedings}, Vol.\ \bibinfo {volume} {8071},\ \bibinfo {editor} {edited by\
  \bibinfo {editor} {\bibfnamefont {M.}~\bibnamefont {Bertolotti}}}\ (\bibinfo
  {publisher} {SPIE, Bellingham},\ \bibinfo {year} {2011})\ p.\ \bibinfo
  {pages} {807105}\BibitemShut {NoStop}%
\bibitem [{\citenamefont {Bloembergen}\ and\ \citenamefont
  {Pershan}(1962)}]{Blombergen1962}%
  \BibitemOpen
  \bibfield  {author} {\bibinfo {author} {\bibfnamefont {N.}~\bibnamefont
  {Bloembergen}}\ and\ \bibinfo {author} {\bibfnamefont {P.~S.}\ \bibnamefont
  {Pershan}},\ }\bibfield  {title} {\enquote {\bibinfo {title} {Light waves at
  the boundary of nonlinear media},}\ }\href {\doibase 10.1103/PhysRev.128.606}
  {\bibfield  {journal} {\bibinfo  {journal} {Phys. Rev.}\ }\textbf {\bibinfo
  {volume} {128}},\ \bibinfo {pages} {606--622} (\bibinfo {year}
  {1962})}\BibitemShut {NoStop}%
\bibitem [{\citenamefont {Bloembergen}\ \emph {et~al.}(1969)\citenamefont
  {Bloembergen}, \citenamefont {Simon},\ and\ \citenamefont
  {Lee}}]{Blombergen1969}%
  \BibitemOpen
  \bibfield  {author} {\bibinfo {author} {\bibfnamefont {N.}~\bibnamefont
  {Bloembergen}}, \bibinfo {author} {\bibfnamefont {H.~J.}\ \bibnamefont
  {Simon}}, \ and\ \bibinfo {author} {\bibfnamefont {C.~H.}\ \bibnamefont
  {Lee}},\ }\bibfield  {title} {\enquote {\bibinfo {title} {Total reflection
  phenomena in second-harmonic generation of light},}\ }\href {\doibase
  10.1103/PhysRev.181.1261} {\bibfield  {journal} {\bibinfo  {journal} {Phys.
  Rev.}\ }\textbf {\bibinfo {volume} {181}},\ \bibinfo {pages} {1261--1271}
  (\bibinfo {year} {1969})}\BibitemShut {NoStop}%
\bibitem [{\citenamefont {Mlejnek}\ \emph {et~al.}(1999)\citenamefont
  {Mlejnek}, \citenamefont {Wright}, \citenamefont {Moloney},\ and\
  \citenamefont {Bloembergen}}]{Mlejnek1999}%
  \BibitemOpen
  \bibfield  {author} {\bibinfo {author} {\bibfnamefont {M.}~\bibnamefont
  {Mlejnek}}, \bibinfo {author} {\bibfnamefont {E.~M.}\ \bibnamefont {Wright}},
  \bibinfo {author} {\bibfnamefont {J.~V.}\ \bibnamefont {Moloney}}, \ and\
  \bibinfo {author} {\bibfnamefont {N.}~\bibnamefont {Bloembergen}},\
  }\bibfield  {title} {\enquote {\bibinfo {title} {Second harmonic generation
  of femtosecond pulses at the boundary of a nonlinear dielectric},}\
  }\href@noop {} {\bibfield  {journal} {\bibinfo  {journal} {Phys. Rev. Lett.}\
  }\textbf {\bibinfo {volume} {83}},\ \bibinfo {pages} {2934--2937} (\bibinfo
  {year} {1999})}\BibitemShut {NoStop}%
\bibitem [{\citenamefont {Centini}\ \emph {et~al.}(2008)\citenamefont
  {Centini}, \citenamefont {Roppo}, \citenamefont {Fazio}, \citenamefont
  {Pettazzi}, \citenamefont {Sibilia}, \citenamefont {Haus}, \citenamefont
  {Foreman}, \citenamefont {Akozbek}, \citenamefont {Bloemer},\ and\
  \citenamefont {Scalora}}]{Centini2008}%
  \BibitemOpen
  \bibfield  {author} {\bibinfo {author} {\bibfnamefont {M.}~\bibnamefont
  {Centini}}, \bibinfo {author} {\bibfnamefont {V.}~\bibnamefont {Roppo}},
  \bibinfo {author} {\bibfnamefont {E.}~\bibnamefont {Fazio}}, \bibinfo
  {author} {\bibfnamefont {F.}~\bibnamefont {Pettazzi}}, \bibinfo {author}
  {\bibfnamefont {C.}~\bibnamefont {Sibilia}}, \bibinfo {author} {\bibfnamefont
  {J.~W.}\ \bibnamefont {Haus}}, \bibinfo {author} {\bibfnamefont {J.~V.}\
  \bibnamefont {Foreman}}, \bibinfo {author} {\bibfnamefont {N.}~\bibnamefont
  {Akozbek}}, \bibinfo {author} {\bibfnamefont {M.~J.}\ \bibnamefont
  {Bloemer}}, \ and\ \bibinfo {author} {\bibfnamefont {M.}~\bibnamefont
  {Scalora}},\ }\bibfield  {title} {\enquote {\bibinfo {title} {Inhibition of
  linear absorption in opaque materials using phase-locked harmonic
  generation},}\ }\href@noop {} {\bibfield  {journal} {\bibinfo  {journal}
  {Phys. Rev. Lett.}\ }\textbf {\bibinfo {volume} {101}},\ \bibinfo {pages}
  {113905} (\bibinfo {year} {2008})}\BibitemShut {NoStop}%
\bibitem [{\citenamefont {{Pe\v{r}ina~Jr.}}\ \emph
  {et~al.}(2009{\natexlab{c}})\citenamefont {{Pe\v{r}ina~Jr.}}, \citenamefont
  {Luk\v{s}},\ and\ \citenamefont {Haderka}}]{PerinaJr2009a}%
  \BibitemOpen
  \bibfield  {author} {\bibinfo {author} {\bibfnamefont {J.}~\bibnamefont
  {{Pe\v{r}ina~Jr.}}}, \bibinfo {author} {\bibfnamefont {A.}~\bibnamefont
  {Luk\v{s}}}, \ and\ \bibinfo {author} {\bibfnamefont {O.}~\bibnamefont
  {Haderka}},\ }\bibfield  {title} {\enquote {\bibinfo {title} {Emission of
  photon pairs at discontinuities of nonlinearity},}\ }\href@noop {} {\bibfield
   {journal} {\bibinfo  {journal} {Phys. Rev. A}\ }\textbf {\bibinfo {volume}
  {80}},\ \bibinfo {pages} {043837} (\bibinfo {year}
  {2009}{\natexlab{c}})}\BibitemShut {NoStop}%
\bibitem [{\citenamefont {Jav\r{u}rek}\ \emph
  {et~al.}(2014{\natexlab{b}})\citenamefont {Jav\r{u}rek}, \citenamefont
  {Svozil\'{i}k},\ and\ \citenamefont {Pe\v{r}ina}}]{Javurek2014c}%
  \BibitemOpen
  \bibfield  {author} {\bibinfo {author} {\bibfnamefont {D.}~\bibnamefont
  {Jav\r{u}rek}}, \bibinfo {author} {\bibfnamefont {J.}~\bibnamefont
  {Svozil\'{i}k}}, \ and\ \bibinfo {author} {\bibfnamefont {J.}~\bibnamefont
  {Pe\v{r}ina}},\ }\bibfield  {title} {\enquote {\bibinfo {title} {Photon-pair
  generation in nonlinear metal-dielectric one-dimensional photonic
  structures},}\ }\href {\doibase 10.1103/PhysRevA.90.053813} {\bibfield
  {journal} {\bibinfo  {journal} {Phys. Rev. A}\ }\textbf {\bibinfo {volume}
  {90}},\ \bibinfo {pages} {053813} (\bibinfo {year}
  {2014}{\natexlab{b}})}\BibitemShut {NoStop}%
\bibitem [{\citenamefont {Huttner}\ \emph {et~al.}(1990)\citenamefont
  {Huttner}, \citenamefont {Serulnik},\ and\ \citenamefont
  {Ben-Aryeh}}]{Huttner1990}%
  \BibitemOpen
  \bibfield  {author} {\bibinfo {author} {\bibfnamefont {B.}~\bibnamefont
  {Huttner}}, \bibinfo {author} {\bibfnamefont {S.}~\bibnamefont {Serulnik}}, \
  and\ \bibinfo {author} {\bibfnamefont {Y.}~\bibnamefont {Ben-Aryeh}},\
  }\bibfield  {title} {\enquote {\bibinfo {title} {Quantum analysis of light
  propagation in a parametric amplifier},}\ }\href@noop {} {\bibfield
  {journal} {\bibinfo  {journal} {Phys. Rev. A}\ }\textbf {\bibinfo {volume}
  {42}},\ \bibinfo {pages} {5594---5600} (\bibinfo {year} {1990})}\BibitemShut
  {NoStop}%
\bibitem [{\citenamefont {Ben-Aryeh}\ and\ \citenamefont
  {Serulnik}(1991)}]{BenAryeh1991}%
  \BibitemOpen
  \bibfield  {author} {\bibinfo {author} {\bibfnamefont {Y.}~\bibnamefont
  {Ben-Aryeh}}\ and\ \bibinfo {author} {\bibfnamefont {S.}~\bibnamefont
  {Serulnik}},\ }\bibfield  {title} {\enquote {\bibinfo {title} {The quantum
  treatment of propagation in non-linear optical media by the use of temporal
  modes},}\ }\href {\doibase http://dx.doi.org/10.1016/0375-9601(91)90650-W}
  {\bibfield  {journal} {\bibinfo  {journal} {Phys. Lett. A}\ }\textbf
  {\bibinfo {volume} {155}},\ \bibinfo {pages} {473 -- 479} (\bibinfo {year}
  {1991})}\BibitemShut {NoStop}%
\bibitem [{\citenamefont {Luk\v{s}}\ \emph {et~al.}(2012)\citenamefont
  {Luk\v{s}}, \citenamefont {Pe\v{r}inov\'a},\ and\ \citenamefont
  {K\v{r}epelka}}]{Luks2012}%
  \BibitemOpen
  \bibfield  {author} {\bibinfo {author} {\bibfnamefont {A.}~\bibnamefont
  {Luk\v{s}}}, \bibinfo {author} {\bibfnamefont {V.}~\bibnamefont
  {Pe\v{r}inov\'a}}, \ and\ \bibinfo {author} {\bibfnamefont {J.}~\bibnamefont
  {K\v{r}epelka}},\ }\bibfield  {title} {\enquote {\bibinfo {title} {{Surface
  effect on spontaneous parametric down-conversion}},}\ }in\ \href@noop {}
  {\emph {\bibinfo {booktitle} {Wave and Quantum Aspects of Contemporary
  Optics}}},\ \bibinfo {series} {SPIE Conference proceedings}, Vol.\ \bibinfo
  {volume} {8697},\ \bibinfo {editor} {edited by\ \bibinfo {editor}
  {\bibfnamefont {J.}~\bibnamefont {{Pe\v{r}ina~Jr.}}}, \bibinfo {editor}
  {\bibfnamefont {L.}~\bibnamefont {No\v{z}ka}}, \bibinfo {editor}
  {\bibfnamefont {M.}~\bibnamefont {Hrabovsk\'y}}, \bibinfo {editor}
  {\bibfnamefont {D.}~\bibnamefont {Sender\'akov\'a}}, \ and\ \bibinfo {editor}
  {\bibfnamefont {W.}~\bibnamefont {Urba\'nczyk}}}\ (\bibinfo  {publisher}
  {SPIE, Bellingham},\ \bibinfo {year} {2012})\ p.\ \bibinfo {pages} {UNSP
  869726}\BibitemShut {NoStop}%
\bibitem [{\citenamefont {Pe\v{r}inov\'{a}}\ \emph {et~al.}(2013)\citenamefont
  {Pe\v{r}inov\'{a}}, \citenamefont {Luk\v{s}},\ and\ \citenamefont
  {{Pe\v{r}ina~Jr.}}}]{Perinova2013}%
  \BibitemOpen
  \bibfield  {author} {\bibinfo {author} {\bibfnamefont {V.}~\bibnamefont
  {Pe\v{r}inov\'{a}}}, \bibinfo {author} {\bibfnamefont {A.}~\bibnamefont
  {Luk\v{s}}}, \ and\ \bibinfo {author} {\bibfnamefont {J.}~\bibnamefont
  {{Pe\v{r}ina~Jr.}}},\ }\bibfield  {title} {\enquote {\bibinfo {title}
  {Quantization of radiation emitted at discontinuities of nonlinearity},}\
  }\href@noop {} {\bibfield  {journal} {\bibinfo  {journal} {Phys. Scr.}\
  }\textbf {\bibinfo {volume} {T153}},\ \bibinfo {pages} {014050} (\bibinfo
  {year} {2013})}\BibitemShut {NoStop}%
\bibitem [{\citenamefont {Yeh}(1988)}]{Yeh1988}%
  \BibitemOpen
  \bibfield  {author} {\bibinfo {author} {\bibfnamefont {P.}~\bibnamefont
  {Yeh}},\ }\href@noop {} {\emph {\bibinfo {title} {Optical Waves in Layered
  Media}}}\ (\bibinfo  {publisher} {Wiley, New York},\ \bibinfo {year}
  {1988})\BibitemShut {NoStop}%
\end{thebibliography}

%

\end{document}